\documentstyle[aps,eqsecnum,preprint,multicol]{revtex}

\begin{document}
\title{\bf Grand canonical ensemble simulation studies of polydisperse fluids}
\author{Nigel B. Wilding}
\address{Department of Mathematical Sciences, The University of Liverpool,\\
Liverpool L69 7ZL, U.K.}

\author{Peter Sollich} 
\address{Department of Mathematics, King's College, University of
London,\\ Strand, London WC2R 2LS, UK}

\input epsf
\epsfclipon

\tighten
\maketitle

\begin{abstract} 

We describe a Monte Carlo scheme for simulating polydisperse fluids
within the grand canonical ensemble. Given some polydisperse attribute
$\sigma$, the state of the system is described by a density
distribution $\rho(\sigma)$ whose form is controlled by the imposed
chemical potential distribution $\mu(\sigma)$. We detail how histogram
extrapolation techniques can be employed to tune $\mu(\sigma)$ such as
to traverse some particular desired path in the space of
$\rho(\sigma)$. The method is applied in simulations of size-disperse
hard spheres with densities distributed according to Schulz and
log-normal forms. In each case, the equation of state is obtained along
the dilution line, i.e. the path along which the scale of
$\rho(\sigma)$ changes but not its shape. The results are compared with
the moment-based expressions of Boublik et al (J. Chem. Phys.  {\bf
54}, 1523 (1971)) and Salacuse and Stell (J. Chem. Phys. {\bf 77}, 3714
(1982)).  It is found that for high degrees of polydispersity, both
expressions fail to give a quantitatively accurate description of the
equation of state when the overall volume fraction is large.

\end{abstract}




\section{Introduction}
\label{sec:intro}

Statistical mechanics was originally formulated to describe the
properties of systems of {\em identical} particles such as atoms or
small molecules. However, many materials of industrial and commercial
importance do not fit neatly into this framework. For example, the
particles in a colloidal suspension are never precisely identical to one
another, but have a range of radii (and possibly surface charges, shapes
etc). This dependence of the particle properties on one or more
continuous parameters is known as polydispersity. 

To process a polydisperse colloidal material, one needs to know its
phase behavior, i.e. the conditions of temperature and pressure under
which a given structure is thermodynamically stable. The main obstacles
to gaining this information arise from the effectively infinite number
of particle species present in a polydisperse system. Labeling these
by the continuous polydispersity attribute $\sigma$, the state of the
system must be described by a density distribution $\rho(\sigma)$,
rather than a finite number of density variables. The phase diagram is
therefore infinite dimensional, a feature that poses serious problems
to experiment and theory alike.

The chief difficulty faced in {\em experimental} studies of
polydisperse systems is that the infinite dimensionality of the phase
diagram precludes a complete mapping of the phase behavior. Instead one
is forced to focus attention on particular low dimensional manifolds
(slices) of the full diagram. Typically this involves determining the
system properties along some desired trajectory through the space of
$\rho(\sigma)$. Such a strategy is often pursued in experiments on
colloidal suspensions \cite{COLLOIDS,BAUS}, where the phase behavior is
studied along a so-called {\em dilution line}. The experimental
procedure for tracking this line involves adding a prescribed quantity
of colloid of some known degree of polydispersity to a vessel of fixed
volume $V$, the remaining volume being occupied by a solvent. The
number distribution of colloidal particles $N(\sigma)$ determines the
density distribution of the suspension, $\rho(\sigma)\equiv
N(\sigma)/V$. Since in a given substance, the relative proportions of
the number of particles of each $\sigma$ are fixed, changing the amount
of colloid added simply alters the {\em scale} of $\rho(\sigma)$, not
its {\em shape}. Thus, by varying $N(\sigma)$ at fixed $V$ (or vice
versa) one traces out a locus in the phase diagram in which only the
overall scale of $\rho(\sigma)$ changes.

As regards {\em theoretical} studies of phase behavior, these typically
endeavor to calculate the system free energy as a function of a set of
density variables. The difficulty in achieving this for a polydisperse
system is that the free energy $f[\rho(\sigma)]$ is a functional of
$\rho(\sigma)$, and therefore itself occupies an infinite dimensional
space. This renders intractable the task of identifying phase
boundaries. Recently, however, progress has stemmed from the
observation that it is possible to approximate the full free energy by
a so-called ``moment free energy'' containing the full ideal gas
contribution plus an excess part that depends only on certain principal
moments of the full excess free energy \cite{SOLLICH}. Doing so reduces
the problem to a finite number of density variables and allows
calculation of phase coexistence properties within a systematically
refinable approximation scheme. Additionally the theory delivers (for
the given free energy) exact results for the location of
spinodals, critical points and the cloud and shadow curves. Use of this
approach promises to enhance significantly our understanding of phase
behavior in polydisperse systems.

In view of the approximations inherent in theoretical approaches, it is
natural to consider deploying {\rm computer simulation} to study the
phase behavior of polydisperse colloidal fluids. Although simulations
(like experiment) are restricted to studying limited regions of the
phase diagram such as the dilution line, they have the advantage that
can be used to investigate the {\em same} model systems as studied
theoretically.  Furthermore, they deliver (modulo finite-size effects)
essentially exact results, providing invaluable benchmarks against
which to test theoretical predictions. Sometimes too, the physical
insight gleaned from simulations serves as the impetus for fresh
theoretical advances.

One simulation approach for obtaining the thermodynamic properties of a
polydisperse system is to simply mimic the experimental procedure. This
can be achieved by employing a canonical ensemble (CE) formalism
wherein a simulation box of fixed volume $V$ is populated by a
prescribed number of particles $N$, whose sizes are distributed
according to the desired $N(\sigma)$. It practice, however, it
transpires that the CE represents a far from optimal framework for
simulating polydisperse fluids. The principal difficulty lies with the
limited range of computationally accessible particle numbers which, in
any simulation, is typically many orders of magnitude smaller than
found in an experiment. The resulting finite-size effects are
particularly pronounced in a CE simulation because the specific
realization of the disorder $N(\sigma)$ is {\em fixed}. This suppresses
large scale fluctuations in $\rho(\sigma)$ and potentially leads to
sampling deficiencies \cite{FOOT1}. Additionally, the CE suffers other
drawbacks familiar from simulation studies of monodisperse fluids. For
instance, relaxation times are extended because density fluctuations
decay solely via diffusion; there is no direct access to information on
chemical potentials; metastability and hysteresis hinder the study of
phase transitions.

Experience with the simulation of monodisperse fluids has shown that use
of the grand canonical ensemble (GCE) is highly effective in
circumventing many of the aforementioned problems associated with the CE
\cite{WILDING95,PANAGIO00,WILDING01}. As we shall show, its application
in the context of polydisperse fluids retains many of the benefits of
the monodisperse case. Moreover, it provides the key to improved
sampling of the density distribution $\rho(\sigma)$. This is because
within the GCE framework $N(\sigma)$ fluctuates {\em as a whole},
thereby capturing fluctuations in $\rho(\sigma)$ on all simulation
length scales.  Notwithstanding these advantages, however, the GCE might
appear, at first sight, unsuitable for the purpose of traversing a {\em
particular} trajectory through the space of the density distribution
$\rho(\sigma)$. This is because $\rho(\sigma)$ ostensibly lies out-with
the direct control of the simulator, its form being instead determined
by the imposed chemical potential distribution $\mu(\sigma)$.
Nevertheless, it turns out to be possible to {\em tune} $\mu(\sigma)$
within a histogram extrapolation scheme, in such a way as to realize a
specific desired form of $\rho(\sigma)$.  We shall demonstrate that this
dual use of the GCE and histogram extrapolation methods permits a chosen
phase space path to be followed efficiently.

The layout of our paper is as follows. In sec.~\ref{sec:statmech} we
formulate the statistical mechanics for a polydisperse fluid within the
grand canonical ensemble. We then describe (sec.~\ref{sec:implement})
the combined GCE plus histogram extrapolation methodology for tracking a
particular path through the space of $\rho(\sigma)$. In
section~\ref{sec:dilution} we apply the method to the problem of
obtaining the dilution line properties of three types of size-disperse
hard sphere fluids.  The chemical potential distribution of these
fluids is determined as a function of volume fraction and the results
compared with the predictions of two commonly used equations of state.
Finally in section~\ref{sec:conclusions}, we discuss our findings and
their implications.

\section{Method}

\subsection{Statistical Mechanics}
\label{sec:statmech}

We consider a classical fluid of polydisperse particles confined to a
volume $V=L^d$. The system is assumed to be thermodynamically open, so
that the particle-number distribution $N(\sigma)$ is a statistical
quantity. The associated grand canonical partition function takes the
form:

\begin{equation} 
\label{eq:bigzdef} 
{\cal Z}_V  = \sum _{N=0}^{\infty }\frac{1}{N!}\prod_{i=1}^{N} \left\{\int_V d\vec{r}_i\int_0^\infty d\sigma_i\right\} \exp{\left(-\beta{\cal H}_N\left(\{\vec{r},\sigma\}\right)\right)}
\end{equation} 
with
\begin{equation}
{\cal H}_N\left(\{\vec{r},\sigma\}\right)=\Phi \left(\{ \vec{r},\sigma \}\right)-\sum_{i=1}^N\mu(\sigma_i)\;.
\end{equation}
Here $N$ is the overall particle number, while $\beta=(k_BT)^{-1}$  and
$\mu(\sigma)$ are respectively the prescribed inverse temperature and
chemical potential distribution. $\{\vec{r},\sigma\}$ denotes the {\em
configuration}, i.e. the complete set $(\vec{r_1},\sigma_1),(\vec{r_2},
\sigma_2)\cdots (\vec{r_N},\sigma_N)$ of particle position vectors and
polydisperse attributes. The corresponding configurational energy $\Phi
\left(\{ \vec{r},\sigma \}\right)$ is assumed to reside in a sum of
pairwise interactions

\begin{equation}
\Phi \left(\{ \vec{r},\sigma \}\right)= \sum_{i<j =1}^N\phi(\mid\vec{r}_i-\vec{r}_j\mid,\sigma_i,\sigma_j)\;,
\end{equation}
where $\phi$ is the pair potential.

The instantaneous number distribution is defined by

\begin{equation}
N(\sigma) \equiv \sum_{i=1}^N \delta(\sigma-\sigma_i)\;,
\end{equation}
with $\sigma$ the continuous polydispersity attribute and $N=\int
N(\sigma)d\sigma$. We shall be concerned with the fluctuations in the
associated density distribution

\begin{equation}
\rho(\sigma)\equiv N(\sigma)/V\;,
\end{equation}
and the configurational energy density

\[
u\equiv \Phi(\{ \vec{r},\sigma \})/V \;.
\]
The statistical behavior of these observables is completely described
by their joint probability distribution \cite{FOOT2}

\begin{equation}
p_V[\rho(\sigma),u]= \left<\delta \left( u -V^{-1}\Phi (\{ \vec{r},\sigma \}) \right)\prod_\sigma\delta \left( \rho(\sigma) - V^{-1}N(\sigma) \right) \right>\;, 
\label{eq:jointdef}
\end{equation}
or more explicitly

\begin{eqnarray}
p_V[\rho(\sigma),u] &=& \frac{1}{{\cal Z}_V}\sum_{N=0}^\infty\frac{1}{N!}\prod _{i=1}^{N} \left\{\int_V d\vec{r}_i\int_0^\infty d\sigma_i\right\} \exp{\left(-\beta{\cal H}_N\left(\{\vec{r},\sigma\}\right)\right )} \\ \nonumber
&\;& \times\;  \delta \left( u - V^{-1}\Phi(\{ \vec{r},\sigma \}) \right) \prod_\sigma\delta \left( \rho(\sigma) - V^{-1}N(\sigma) \right) \;.
\label{eq:jointdist}
\end{eqnarray}
Integrating over the energy fluctuations yields the probability
distribution function of the density distribution

\begin{equation}
p_V[\rho(\sigma)]=\int p_V[\rho(\sigma),u]du\;.
\end{equation}
Our specific concern is with the average form of $\rho(\sigma)$, given by

\begin{equation}
\overline{\rho}(\sigma)=\int\rho(\sigma)p_V[\rho(\sigma)]d\rho(\sigma)\;.
\end{equation}
Given a prescribed chemical potential distribution $\mu(\sigma)$ and
temperature $\beta$, the form of $\overline{\rho}(\sigma)$ can be
determined by simulation. Except in the ideal gas limit, however,  no
exact relationship between $\overline{\rho}(\sigma)$ and $\mu(\sigma)$
will generally be available. Thus one cannot (from a `bare' GCE
simulation), readily determine that $\mu(\sigma)$ corresponding to a
particular desired {\em target} density distribution $\rho_t(\sigma)$. 
Subject to certain restrictions however, this can be achieved via use
of histogram extrapolation. 

The key idea of histogram extrapolation \cite{FERRENBERG} is that a
measured distribution $p_V[\rho(\sigma),u]$ accumulated at one set of
model parameters $\mu(\sigma),\beta$ can be reweighted to yield
estimates of the distribution appropriate to other parameters
$\mu^\prime(\sigma),\beta^\prime$. In its simplest form the reweighting
is given by

\begin{equation}
p_V^\prime[\rho(\sigma),u|\mu^\prime(\sigma),\beta^\prime]= w p_V[\rho(\sigma),u|\mu(\sigma),\beta]
\end{equation}
where the reweighting factor $w$ takes the form
\begin{equation}
w=\exp\left(\sum_{i=1}^N\left[\beta^\prime\mu^\prime(\sigma_i)-\beta\mu(\sigma_i)\right]-V\left(\beta^\prime-\beta\right)u\right)  
\label{eqn:histwht}
\end{equation}
By tuning the form of $\mu^\prime(\sigma)$ and the value of
$\beta^\prime$ within the reweighting scheme, it is possible to scan the
space of $\overline{\rho}(\sigma)$, thereby ``homing in'' on the target
density distribution. To this end it is expedient to define a {\em cost
function} measuring the deviation of
$\overline{\rho}(\sigma\mid\mu^\prime(\sigma),\beta^\prime)$ 
from the target form

\begin{equation}
\Delta(\mu^\prime(\sigma),\beta^\prime)\equiv\int \mid\overline{\rho}(\sigma)-\rho_t(\sigma)\mid\gamma(\sigma)d\sigma \;.
\label{eq:costfn}
\end{equation}
Here, for numerical convenience, we have incorporated a weight function
$\gamma(\sigma)$ into our definition, the role of which (as described below)
is to ensure that comparable contributions to the cost function arise
from all sampled regions of the $\sigma$-domain. Within this framework,
the task of determining those values of
$\mu^\prime(\sigma),\beta^\prime$ that yield the target distribution
$\rho_t(\sigma)$ reduces to that of functionally minimizing the cost
function $\Delta$ with respect to $\mu^\prime(\sigma),\beta^\prime$.
As we describe in the next section, this is achievable using standard
algorithms.

\subsection{Implementation}
\label{sec:implement}

We have employed Monte Carlo (MC) simulation within the grand canonical
ensemble to study the dilution line properties of systems of
size-disperse hard spheres. This section details the
principal aspects of our simulation and analysis procedure.

\subsubsection{Program, data structure and acquisition}
\label{sec:program}

Our simulated system comprises a variable number of hard spheres
contained within a periodic box of volume $V=L^d$.  The dimensionless
polydisperse variable $\sigma$ was taken to be the particle {\em
diameter} expressed in units of the mean diameter (see
section~\ref{sec:dilution}). An upper bound $\sigma_c$ was placed on the
permitted range of diameters, and the simulation volume was partitioned
into an array of $l^3$ cells each of linear dimension $\sigma_c$, so
that $L=l\sigma_c$. This strategy aids efficient identification of
particle interactions by ensuring that interacting particles occupy
either the same cell or directly neighboring ones. 

The grand canonical ensemble Monte Carlo (GCMC) algorithm employed has a
Metropolis form \cite{FRENKEL} and invokes four types of operation:
particle displacements, particle insertions, particle deletions and
particle resizing; each is attempted with equal frequency. Specific to
the polydisperse case is the resizing operation which entails attempting
to change the diameter of a nominated particle by an amount drawn from a
uniform random deviate constrained to lie in some prescribed range. This
range (maximum diameter step-size) is chosen to provide a suitable
balance between efficient sampling and a satisfactory acceptance rate at
the prevailing number density. As regards the remaining types of moves,
these proceed in a manner similar to the monodisperse case
\cite{FRENKEL}, except that for insertion attempts the new particle
diameter is drawn with uniform probability from the range $\sigma \in
[0,\sigma_c]$. 

As primary {\em input}, the program takes the chemical potential
distribution $\mu(\sigma)$, which is required for the accept/reject
Monte Carlo lottery. This distribution is stored in the form of a
histogram, constructed by dividing the truncated interval
$0<\sigma<\sigma_c$ into a prescribed number $M$ of sub-intervals or
``bins'' \cite{NOTE0}. Thus all particles whose diameters fall within the
scope of a given bin are associated with the same value of the chemical
potential.

The principal {\em observable} of interest is the joint probability
distribution $p_V[\rho(\sigma),u]$ (c.f. Eq.~\ref{eq:jointdist}).
Operationally it is impractical to construct the full form of this
distribution in a simulation because to do so would entail forming a
histogram over histograms--the memory storage costs of which would be
prohibitive. The procedure adopted, therefore, was to sample the {\em
instantaneous} density histogram $\rho(\sigma)$ (discretized in the same
manner as $\mu(\sigma)$), and append successive measurements of this
quantity to a file \cite{NOTE0A}. The set of all such samples
constitutes a {\em list} representing a sequential history of the
individual data measurements \cite{WILDING01}. This data list is
post-processed by an analysis program which reads in each of the
individual list entries and averages over all of them to construct a
histogram approximation to the average density distribution
$\bar\rho(\sigma)$. If desired, the analysis program additionally
implements histogram reweighting of the data in order to enable
extrapolation to neighboring values of the model parameters. This
extrapolation is achieved by assigning a weight to each list entry of
the form given by equation~\ref{eqn:histwht}. The complete set of
weights permits construction of the reweighted histogram.

\subsubsection{Tracking a phase space path}
\label{sec:tracking}

The strategy we have adopted is to traverse the phase space path of
interest in a stepwise fashion, utilizing histogram extrapolation to
proceed from one step to the next. In the general case of particles
having a finite potential, the phase space path may involve changes in
the temperature as well the form of $\rho(\sigma)$. For simplicity of
illustration, however, let us presume that  the path is isothermal (i.e
$\beta={\rm constant}$) and that the form of the chemical potential
distribution, $\mu^{(0)}(\sigma)$ say, is known at some arbitrary point
$\rho^{(0)}(\sigma)$ along the path.  The procedure is then as follows.
From simulations at the known state point, data for
$p_V[\rho(\sigma),u]$ can be accumulated directly. Histogram reweighting
is then applied to this data to extrapolate some distance along the
path to a new point $\rho^{(1)}(\sigma)$, {\em and} to provide an estimate
of the corresponding form of $\mu^{(1)}(\sigma)$. The latter quantity is
then employed in a fresh simulation, the results of which are
extrapolated to a point further along the path, and so on. By iterating
this procedure $\rho^{(i)}(\sigma) \to \rho^{(i+1)}(\sigma)$, $\mu^{(i)}(\sigma)
\to \mu^{(i+1)}(\sigma)$, one traces out the entire phase space path.

The implementation of the extrapolation stage necessitates a prior
choice for the step size, that is the difference between the measured
$\rho(\sigma)$ and the next target $\rho_t(\sigma)$. The magnitude of this
difference should be chosen to be as large as possible, consistent with
remaining within the range of reliable extrapolation. A good indicator
that this is in fact the case is that the individual densities of the
target distribution $\rho_t(\sigma)$ each overlap with the range of
typical fluctuations appearing in the simulation data $\rho(\sigma)$
\cite{FOOT4}. 

Once a suitable step size has been determined, the extrapolation
procedure proceeds by minimizing (within the reweighting scheme) the
cost function $\Delta$ introduced in eq.~\ref{eq:costfn}. For all but
the lowest densities, this task is complicated by the existence of
strong coupling between the $\mu$ variables, deriving from the fact that
the number density for each $\sigma$ depends on the {\em whole} chemical
potential distribution. Fortunately, efficient algorithms for performing
multi-dimensional functional minimization are widely available
\cite{NUMREC} and, at least for the cases we have considered, appear to
operate effectively. The sole difficulty encountered was that, on
occasion, the minimization failed to fully converge for $\sigma$ values
in the wings of $\rho_t(\sigma)$. To remedy this problem, a weight
function $\gamma(\sigma)$ was incorporated in the cost function (c.f.\
eq.~\ref{eq:costfn}), the purpose of which is to enhance the
contribution to $\Delta$ from $\sigma$ values for which $\rho_t(\sigma)$
is small. Good results were obtained by setting $\gamma(\sigma)\propto
[\rho_t(\sigma)]^{-1}$ \cite{FOOT5}. 

It should be stressed that the methodology set out above presumes the
availability of a form of $\mu(\sigma)$ for some starting point on the
phase space path of interest. This may be obtained straightforwardly if
the path passes through a region of low density where reliable
analytical estimates for $\mu(\sigma)$ can be employed. Otherwise the
method must be bootstrapped by other means.  One simple but effective
approach for achieving this operates as follows. Starting from some
initial guess for the desired $\mu(\sigma)$ (e.g.\
$\mu^0(\sigma)=\ln[\rho_t(\sigma)]$) a series of short simulations are
carried out in which $\mu(\sigma)$ is iterated according to 

\begin{equation}
\mu^{(m+1)}(\sigma)=\mu^{(m)}(\sigma)+\left( \frac{\ln\left(\rho_t(\sigma)\right)} {\ln\left(\rho^{(m)}(\sigma)\right)}\right)^\delta\;,
\end{equation}
where $0<\delta<1$ is a damping factor, the value of which may be tuned to optimize convergence.
Although one could certainly envisage more sophisticated and efficient
schemes, we have found this method to operate satisfactorily.

\section{Dilution line studies of polydisperse hard sphere fluids}

\label{sec:dilution}

\subsection{System and simulation details}
\label{sec:simdetails}

We have obtained the dilution line properties (cf. sec.~\ref{sec:intro})
of size-disperse hard sphere \cite{betanote} fluids with diameter distribution
$N(\sigma)$ assigned one of two forms:

\begin{itemize}
\item[(i)]  Schulz distribution.
\item[(ii)] Log-normal distribution.
\end{itemize}
These distributions are conveniently expressed in terms of a normalized
size function $n(\sigma)=N(\sigma)/N$. For the Schulz, this takes the
form

\begin{equation}
n(\sigma)=\frac{1}{z!}\left(\frac{z+1}{\overline{\sigma}}\right)^{z+1}\sigma^z\exp\left[-\left(\frac{z+1}{\overline{\sigma}}\right)\sigma\right]\:,
\label{eq:schulz}
\end{equation}
where $\overline{\sigma}$ is the average particle diameter and $z$ is a
parameter controlling the width of the distribution. For the log-normal
distribution, the size function is given by

\begin{equation}
n(\sigma)=\frac{1+W^2}{\overline{\sigma}\sqrt{2\pi\ln(1+W^2)}}\exp\left(-\frac{[\ln(\sigma/\overline{\sigma})+(3/2)\ln(1+W^2)]^2}{2\ln(1+W^2)}\right)\:,
\label{eq:lognorm}
\end{equation}
with $W$ the standard deviation in units of $\overline{\sigma}$. Note
that both of these distribution are normalized, that is $\int_0^\infty
n(\sigma)d\sigma=1$, and vanish as $\sigma\to 0$, implying a natural
lower limit to $\sigma$. By contrast, there is no finite upper limit
and consequently, for simulation purposes, it was necessary to impose
an upper bound (cut off) $\sigma_c$ (see also
sec.~\ref{sec:implement}).

We have studied {\em three} distinct size distributions---two of the
Schulz form and one of the log-normal form. For the Schulz
distribution, width-parameter values of $z=15$ and $z=5$ were
considered, with cutoff values $\sigma_c=3$ and $\sigma_c=4$
respectively. For the log-normal distribution, the single case $W=2.5$
with $\sigma_c=12$ was studied. In each instance we set
${\overline\sigma}=1$. Cell array sizes (cf. sec.~\ref{sec:program})
of linear dimension $l=3,4,5$ were used. In absolute
dimensionless units ($L=l\sigma_c$) these correspond to $L=9, 12, 15$
respectively for the $z=15$ Schulz, to $L=12, 16, 20$ for the $z=5$
Schulz, and to $L=36, 48, 60$ for the log-normal distribution.  The
histogram discretization parameter (cf. sec.~\ref{sec:implement}) was
set to $M=75$ and $M=100$ for the $z=15$ and $z=5$ Schulz distributions
respectively, and to $M=120$ for the log-normal distribution.

The average density distribution can be expressed in terms of the
normalized size function multiplied by the overall number density
$\rho_0=N/V$, i.e.

\begin{equation} 
\overline{\rho}(\sigma) \equiv \rho_0 n(\sigma)\:.
\end{equation} 
The procedure for tracking the dilution line utilized this overall
density as a measure of the location along the line. The tracking
procedure was initiated at a small value of $\rho_0$ by approximating
the chemical potential distribution according to the ideal gas relation
$\mu(\sigma)=\ln \overline{\rho}(\sigma)$ \cite{FOOT6}. Histogram
extrapolation was applied to the resulting simulation data in order to
refine the initial estimate of $\mu(\sigma)$. Thereafter the strategy
described in sec.~\ref{sec:tracking} was implemented to follow the
dilution line to higher densities. At each step this entailed setting a
target form for $\overline\rho(\sigma)$ corresponding to a value of
$\rho_0$ larger than that used at the previous step. The cost function
measuring the discrepancy between $\overline\rho(\sigma)$ and the target
was then minimized to yield an estimate for the appropriate
$\mu(\sigma)$. On efficiency grounds, this minimization was performed in
two stages; an initial approximation to $\mu(\sigma)$ was obtained from
a one-dimensional minimization in which the activity distribution
$\exp(\mu(\sigma))$ was multiplied by an overall factor. Thereafter the
complete functional minimization was performed to yield a more accurate
form for $\mu(\sigma)$. 

Although (within the specific context of the tracking procedure),
$\rho_0$ provides a convenient measure of the location on the dilution
line, it conveys little information regarding the degree of packing
within a polydisperse system. We shall therefore find it convenient to
quote values for the overall volume fraction of the system given by

\begin{equation}
\eta\equiv\int_0^{\sigma_c}d\sigma \frac{\pi}{6}\sigma^3\bar\rho(\sigma)\:.
\end{equation}
It is this quantity (rather than $\rho_0$) which features in the
presentation of our results, to which we now turn.

\subsection{Results}
\label{sec:results}

Owing to the computational complexity of the simulations, we have
obtained the complete dilution line for each fluid only for the $l=3$
cell array size. For the larger system sizes, only a few spot
measurements were made along the dilution line, each of which was
bootstrapped by appeal to the measured $\mu(\sigma)$ obtained from the
$l=3$ systems.

\begin{figure}[h]
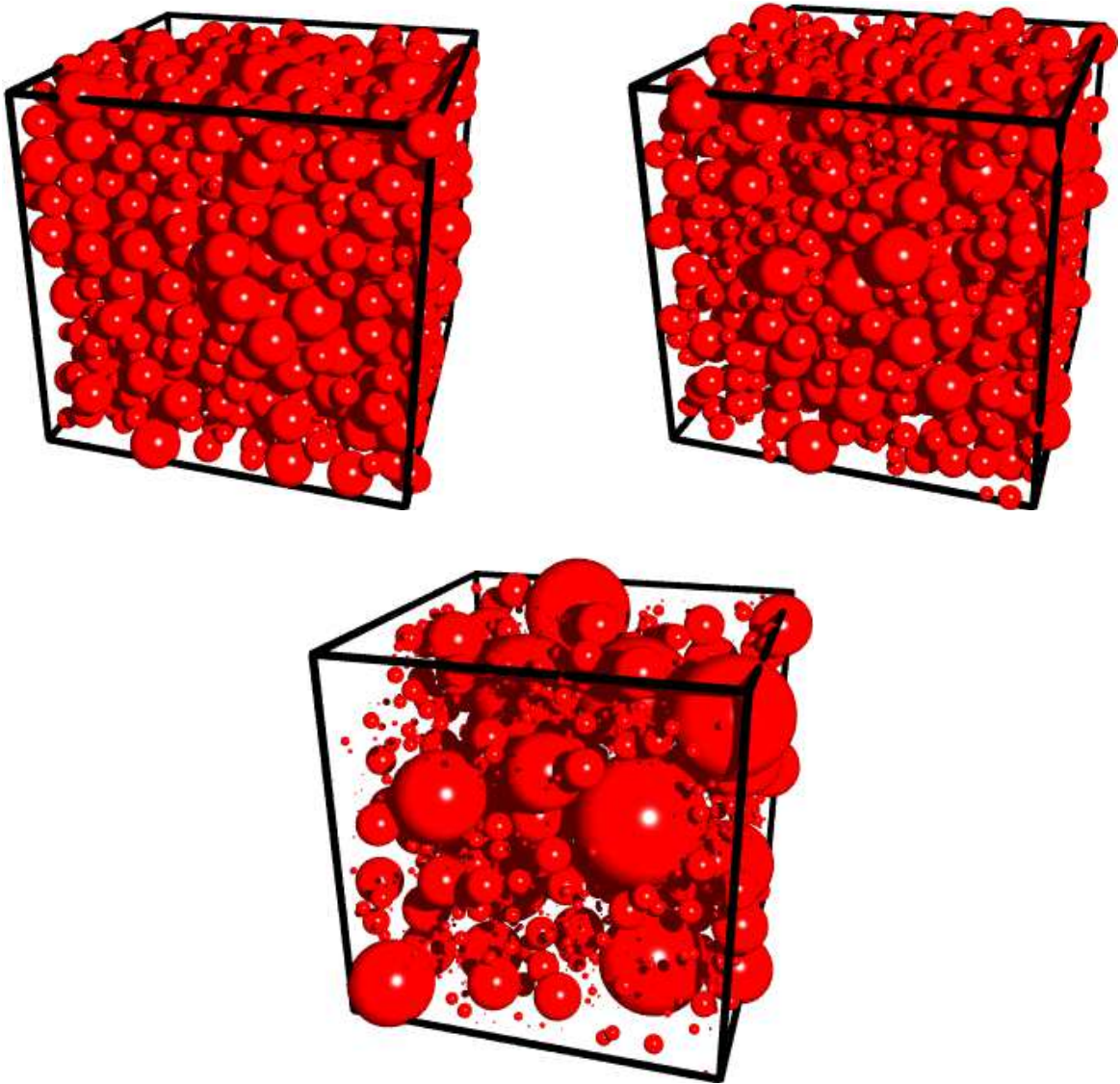

\setlength{\epsfxsize}{6.6cm}
\vspace*{1cm}
\centerline{\mbox{\epsffile{conf_z15_lr.epsi}}\hspace*{2cm}\setlength{\epsfxsize}{6.8cm}\mbox{\epsffile{conf_z5_lr.epsi}}}
\vspace*{0.6cm}
\setlength{\epsfxsize}{7.0cm}
\centerline{\mbox{\epsffile{conf_ln_lr.epsi}}} 
\vspace*{0.5cm}

\caption{Snapshots of configurations. {\bf (a)} Schulz distribution
($z=15$), $\eta=0.43$, $L=12$. {\bf (b)} Schulz distribution ($z=5$),
$\eta=0.38$, $L=16$. {\bf (c)} Log-normal distribution, $\eta=0.29$,
$L=48$.}

\label{fig:snapshots}
\end{figure}

The dilution lines for the three fluids were tracked to the highest
computationally accessible volume fraction, terminating only once the
relaxation timescale became excessively large. The maximum volume
fractions attained were $\eta=0.445$ and $\eta=0.40$ for the Schulz
density distributions with $z=15$ and $z=5$ respectively, and
$\eta=0.33$ for the log-normal case. Snapshot configurations for all
three fluids are shown in fig.~\ref{fig:snapshots} for $\eta$ values
slightly below the maximum attained in each case. We mention in passing
that, at least for the cases of the two Schulz distributions, the
maximum volume fractions reached are somewhat larger than would be
readily attainable for GCMC simulations of monodisperse hard spheres.
The latter, of course, become highly inefficient at large $\eta$
because the acceptance rate for particle insertions falls rapidly as
the free volume diminishes. Whilst the same is true in the
polydisperse context for insertions of large particles, space can often
be found for placing a small particle. This facilitates fluctuations in
the overall particle number $N$, while resizing operations (whose
efficiency at large $\eta$ is typically greater than that of inserting
large particles) ensure continued proper sampling of the density
distribution.

\begin{figure}[h]
\setlength{\epsfxsize}{10.0cm}
\centerline{\mbox{\epsffile{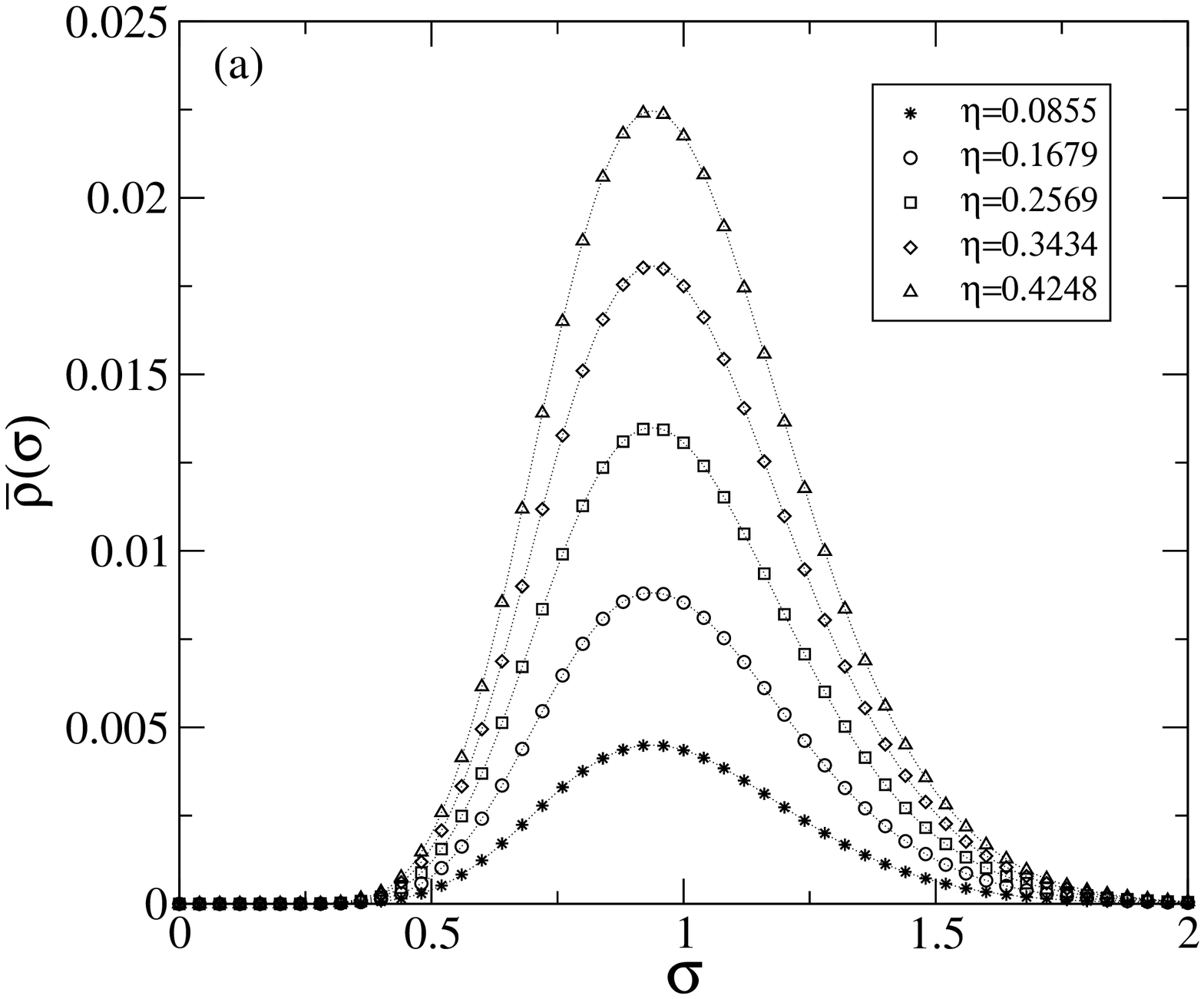}}} 
\setlength{\epsfxsize}{10.0cm}
\centerline{\mbox{\epsffile{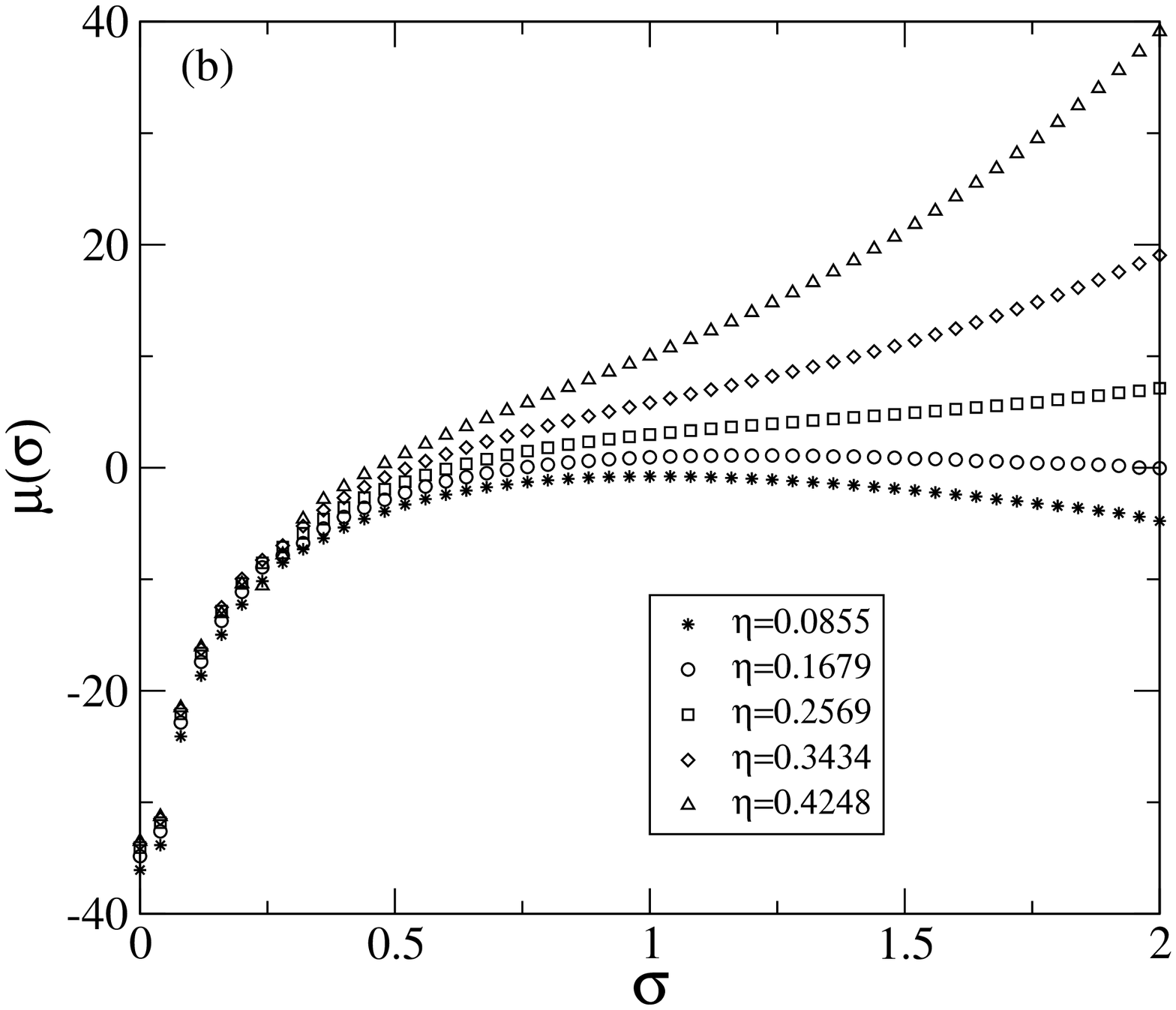}}} 

\caption{Dilution line properties of hard spheres having the Schulz
density distribution ($z=15$, $\sigma_c=3.0$) for system size $L=9$.
{\bf (a)} Data points show the measured density distribution
$\overline{\rho}(\sigma)$ at a selection of values of volume fraction
$\eta$ along the dilution line; dotted lines correspond to the target
distribution $\rho_t(\sigma)$ {\bf (b)} The corresponding  chemical
potential distribution $\mu(\sigma)$. Statistical errors do not exceed
the symbol sizes.}

\label{fig:sz15dists}
\end{figure}

\begin{figure}[h]
\setlength{\epsfxsize}{10.0cm}
\centerline{\mbox{\epsffile{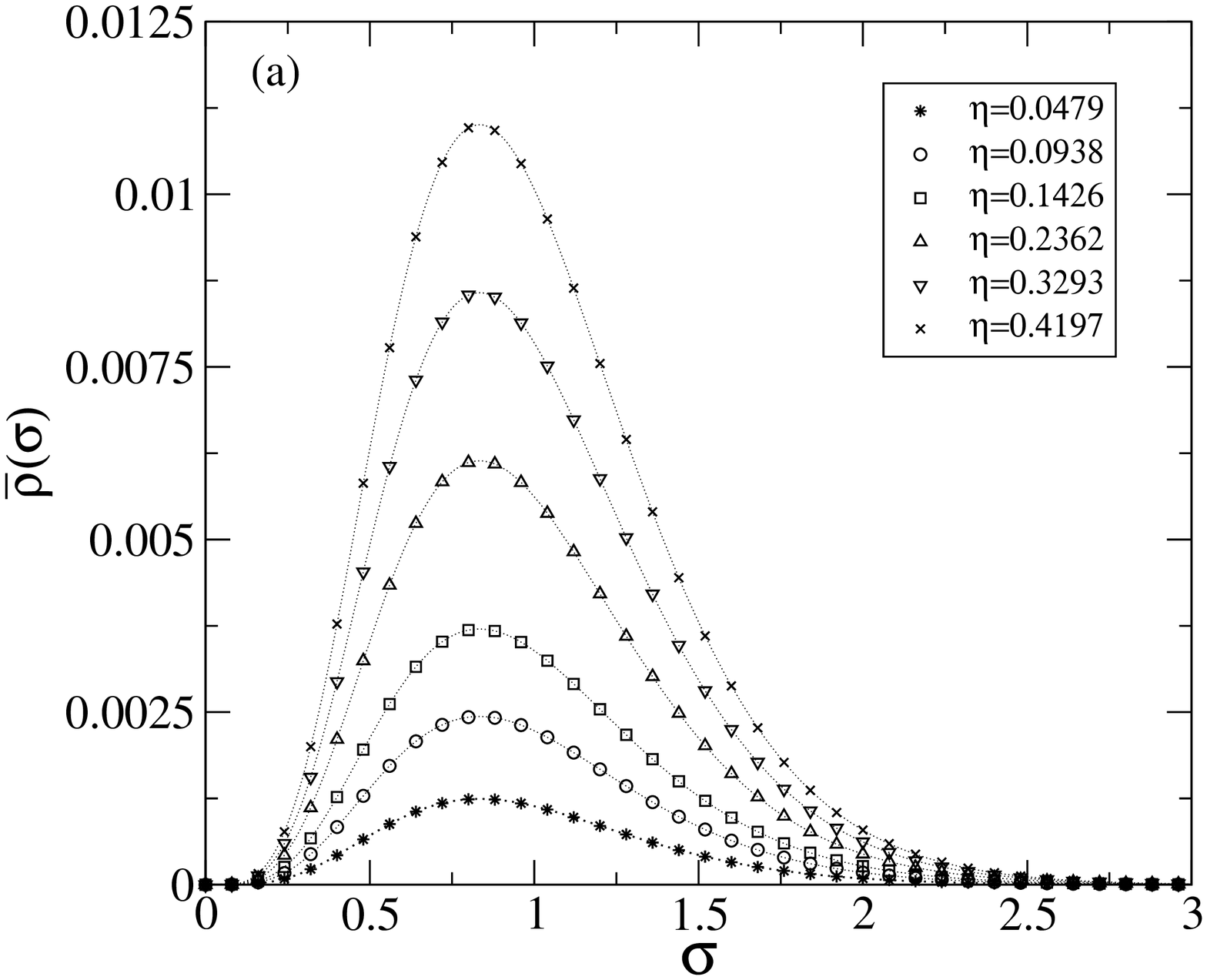}}} 
\setlength{\epsfxsize}{10.0cm}
\vspace*{2cm}
\centerline{\mbox{\epsffile{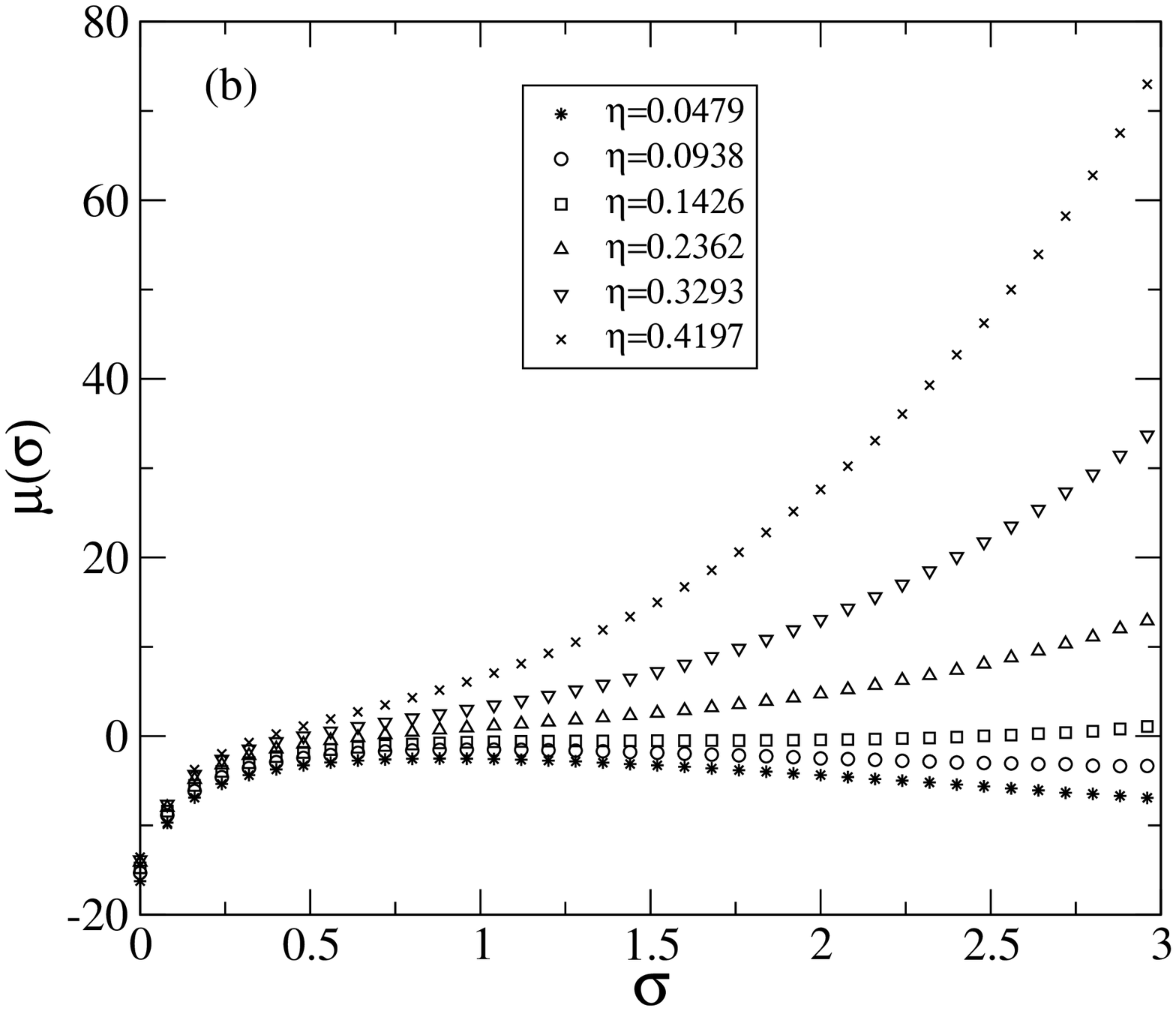}}} 
\vspace*{1cm}

\caption{Dilution line properties of hard spheres having the Schulz
density distribution ($z=5$, $\sigma_c=4$), for system size $L=12$. 
{\bf (a)} Data points show the measured density distribution
$\overline{\rho}(\sigma)$ at a selection of values of volume fraction
$\eta$ along the dilution line; dotted lines correspond to the target
distribution $\rho_t(\sigma)$. {\bf (b)} The corresponding  chemical
potential distribution $\mu(\sigma)$.}

\label{fig:sz5dists}
\end{figure}

The measured forms for $\mu(\sigma)$ as a function of $\eta$  differ
qualitatively between the Schulz and the log-normal density
distributions, and accordingly we discuss them separately. Beginning
with the Schulz case, figs.~\ref{fig:sz15dists} and \ref{fig:sz5dists}
show the measured $\overline{\rho}(\sigma)$ and the corresponding
$\mu(\sigma)$ for the two fluids at a selection of volume fractions
along their respective dilution lines. The range of $\sigma$ values
shown is that for which the simulations delivered data of reasonable
statistical quality.  One notes that for both the $z=15$ and $z=5$
cases, the tails marking the large--$\sigma$ vestiges of the
distributions fall considerably short of the respective cutoffs
$\sigma_c$. Indeed, we have verified that in the course of the
simulations, no particles of diameters approaching the cutoff diameter
occurred on the simulation timescale, implying that our data is
unaffected by its imposition. As regards the forms of the chemical
potential distributions, one sees that for small $\eta$ they display a
maximum near the peak in $\overline{\rho}(\sigma)$---behavior that is
of course mandated at sufficiently low $\eta$ by the known properties
of the ideal gas limit. For larger $\eta$, however, the peak is lost
and $\mu(\sigma)$ increases monotonically. The increase of
$\mu(\sigma)$ in the regime of large $\sigma$ indicates that the excess
chemical potential (measuring the work associated with inserting a
sphere) grows faster with $\sigma$ than the decrease in the ideal
contribution associated with the decay of $\overline{\rho}(\sigma)$.

\begin{figure}[h]
\setlength{\epsfxsize}{10.0cm}
\centerline{\mbox{\epsffile{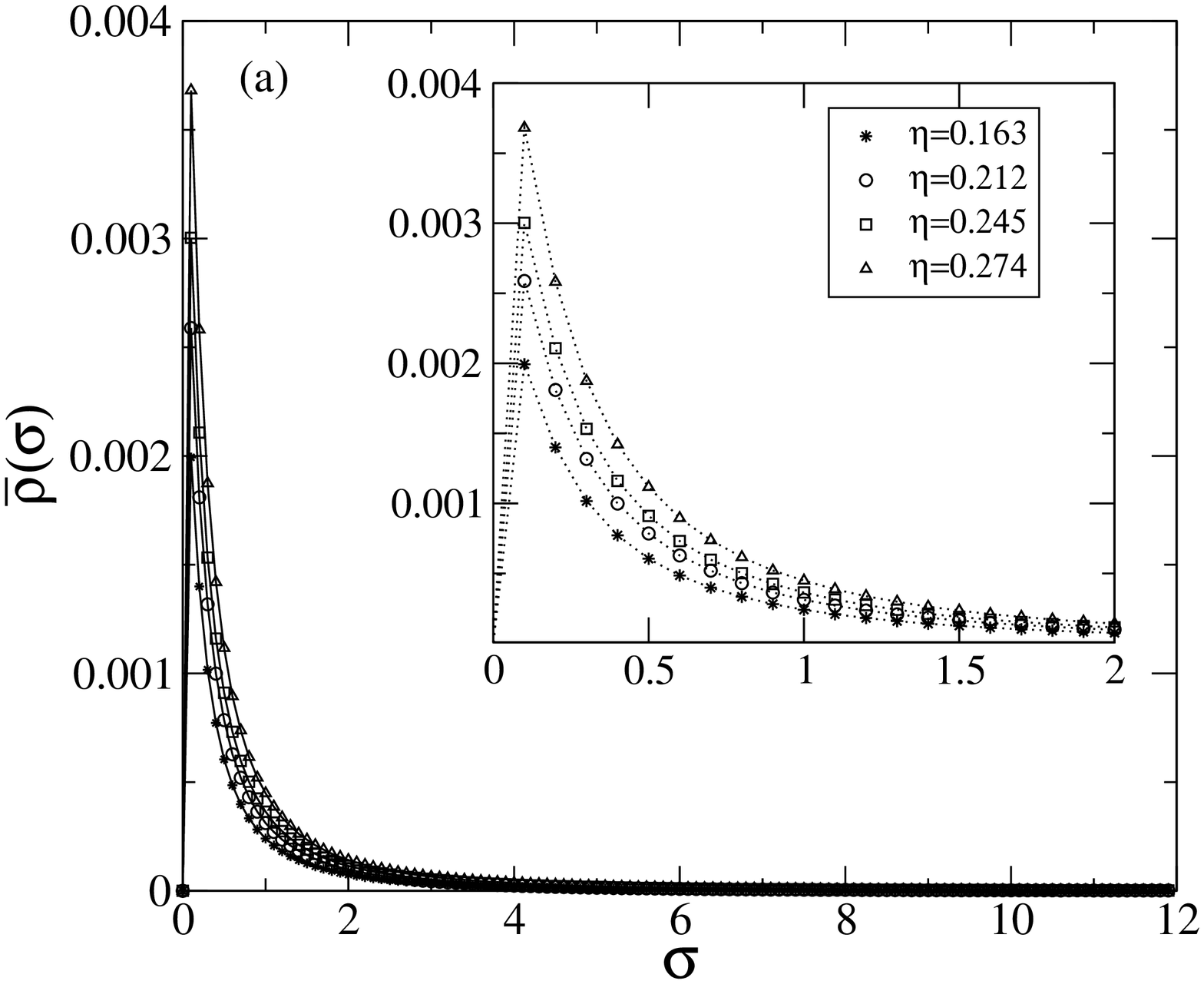}}} 
\setlength{\epsfxsize}{10.0cm}
\vspace*{2cm}
\centerline{\mbox{\epsffile{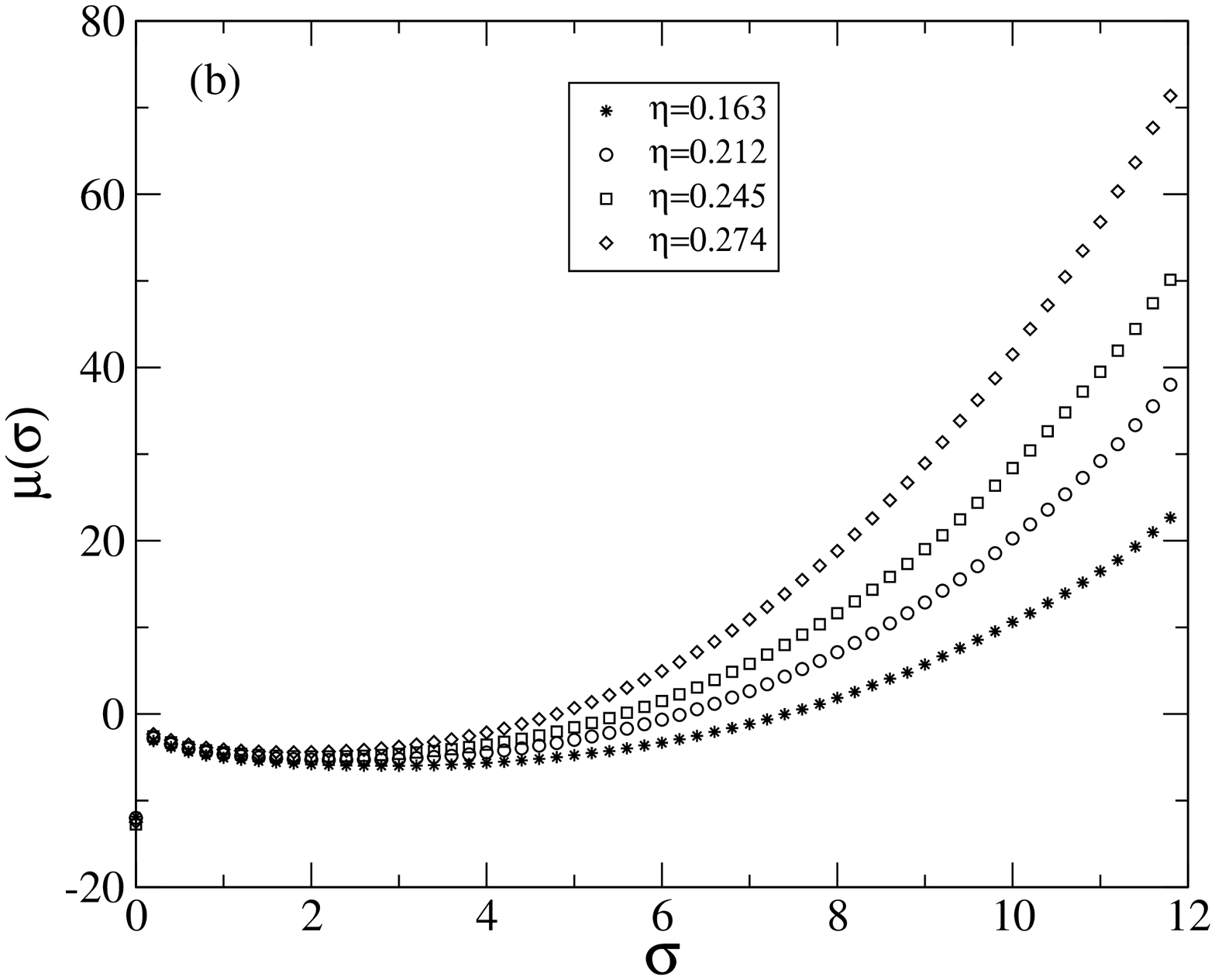}}} 
\vspace*{1cm}

\caption{Dilution line properties of hard spheres having the log-normal
density distribution ($W=2.5$, $\sigma_c=12$) for system size $L=36$. 
{\bf (a)} Data points show the measured density distribution
$\overline{\rho}(\sigma)$ at a selection of values of volume fraction
$\eta$ along the dilution line; dotted lines correspond to the target
distribution $\rho_t(\sigma)$. The inset shows the region of small
$\sigma$. Lines are merely guides to the eye. {\bf (b)} The
corresponding chemical potential distribution $\mu(\sigma)$.
Statistical errors do not exceed the symbol sizes.} 
\label{fig:lndists}
\end{figure}

Turning now to the log-normal case, fig.~\ref{fig:lndists}(a)
demonstrates that $\overline{\rho}(\sigma)$ decays extremely slowly
with increasing $\sigma$. The peak in the distribution therefore occurs
at much smaller $\sigma$ than for the two Schulz forms although the
average diameter ${\overline \sigma}$ is identical in all three cases. As
a practical consequence of the slow decay (and notwithstanding the
imposition of a very large value of $\sigma_c$), ranges of particle
diameters extending up to the cutoff value were observed in the system.
Indeed it was not feasible within the computational constraints to
utilize a cutoff for which this did not occur, and hence truncation
effects are always significant in this system---a point to which we
shall return below. The measured forms of $\mu(\sigma)$ for the
log-normal fluid are shown in fig.~\ref{fig:lndists}(b). In contrast to
the Schulz distributions, they display (for all accessible $\eta$) a
narrow maximum close to the peak in $\overline{\rho}(\sigma)$ at small
$\rho$. Thereafter with increasing $\sigma$ there is a slow fall to a
broad minimum, whereafter $\mu(\sigma)$ increases strongly.

Having outlined the main qualitative features of the relationship
between ${\overline\rho}(\sigma)$ and $\mu(\sigma)$, it is instructive 
to perform a detailed comparison between our measurements and the
predictions of analytical equations of state (EOS) appearing in the
literature. For hard spheres, two commonly used equations are that due
to Boublik, Mansoori, Carnahan, Starling and Leland (BMCSL)
\cite{BOUBLIK,MANSOORI} based on the Carnahan-Starling equation for
monodisperse hard spheres, and that due to Salacuse and Stell
\cite{SALACUSE}, based on the Percus-Yevick (PY) theory. Both are
reproduced in appendix~\ref{sec:append}, and express $\mu(\sigma)$ in
terms of an expansion to third order in $\sigma$, with coefficients
given in terms of the first three moments of ${\overline
\rho}(\sigma)$. We have compared the predictions of the EOS for each of
the three fluids studied, with the finite-size simulation data at three
values of $\eta$, namely a low, a moderate and a high value. We describe
our findings for each fluid in turn.

\begin{figure}[h]
\setlength{\epsfxsize}{7.0cm}
\centerline{\mbox{\epsffile{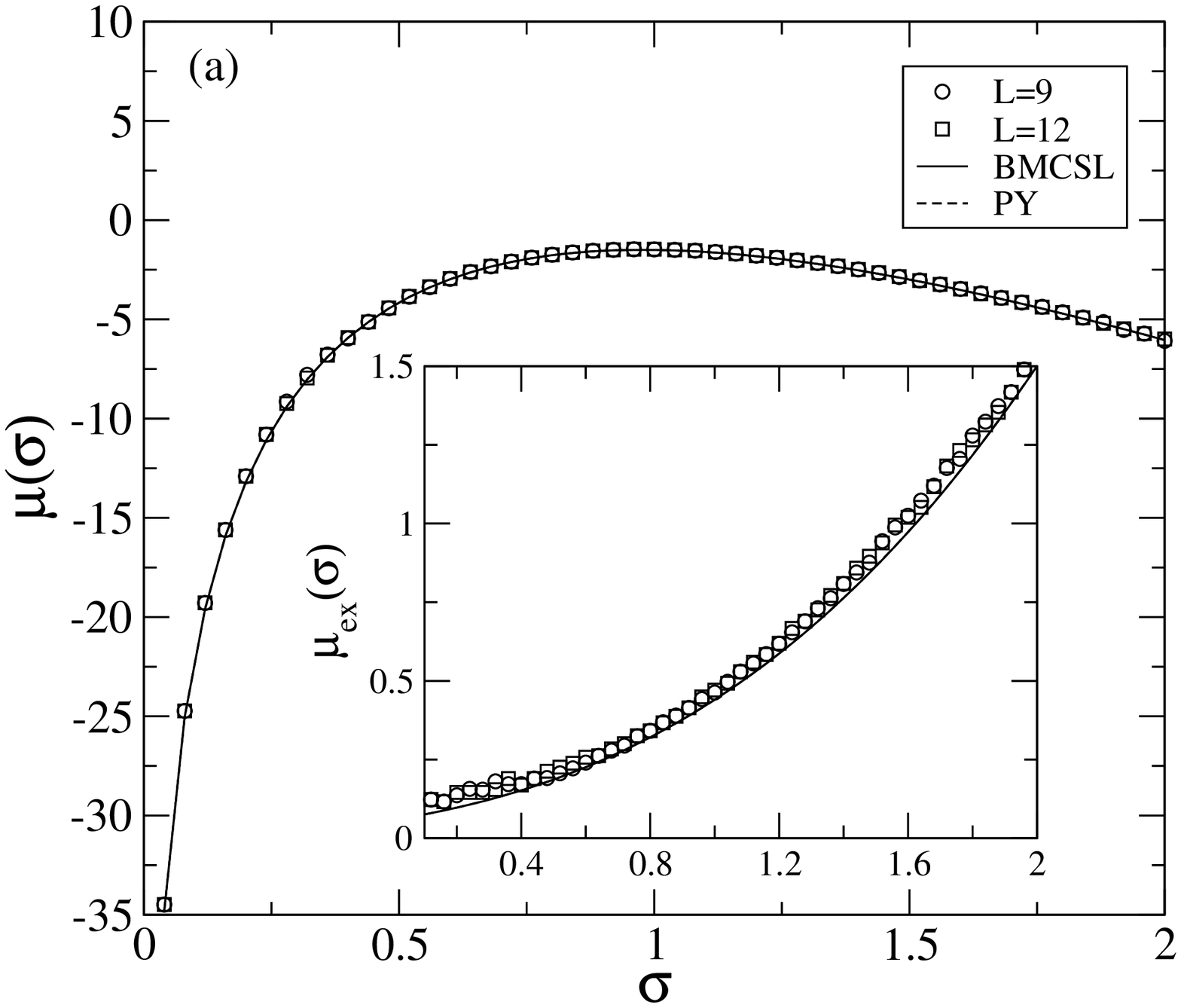}}} 
\vspace*{0.5cm}
\setlength{\epsfxsize}{7.0cm}
\centerline{\mbox{\epsffile{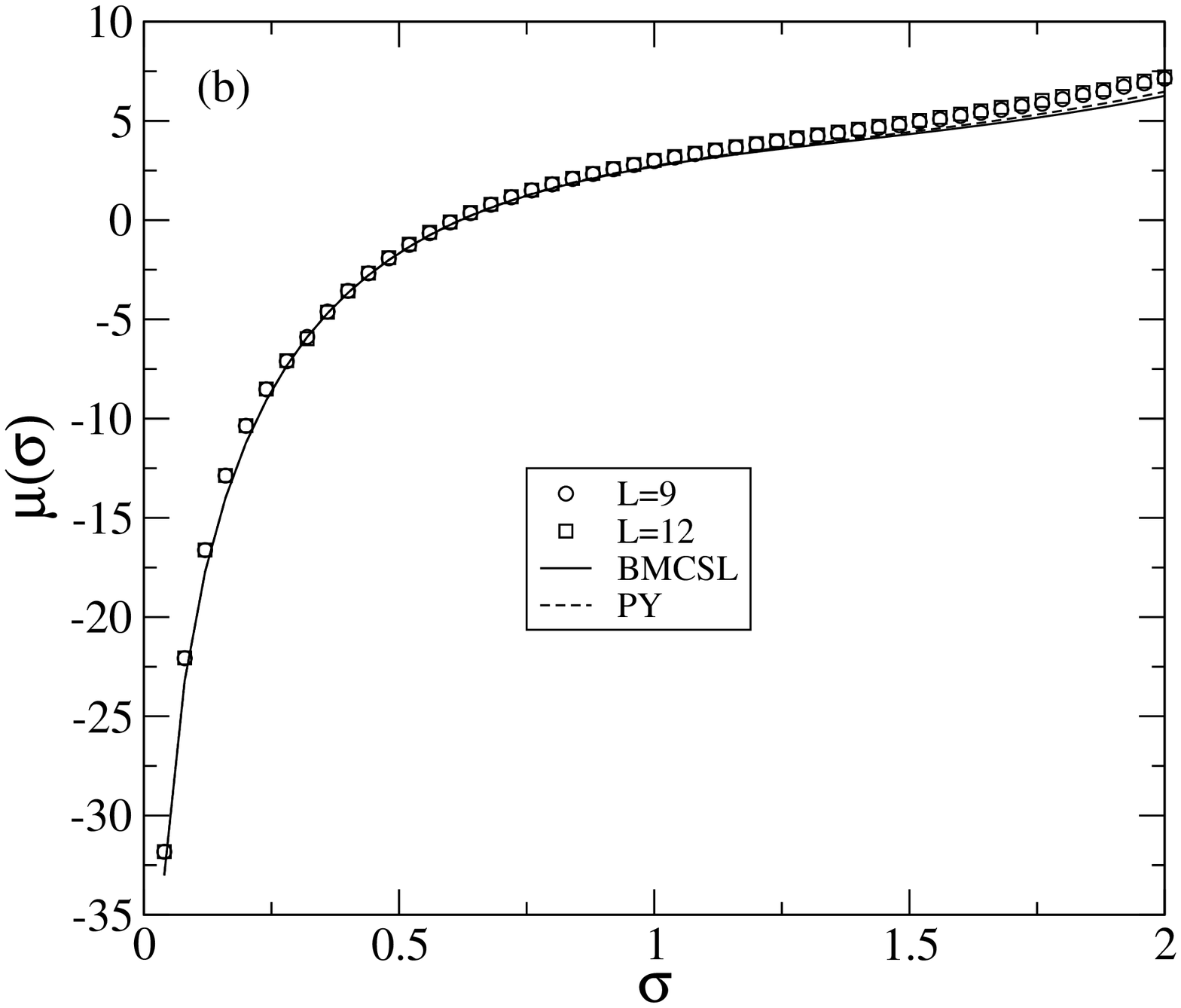}}} 
\setlength{\epsfxsize}{7.0cm}
\vspace*{0.5cm}
\centerline{\mbox{\epsffile{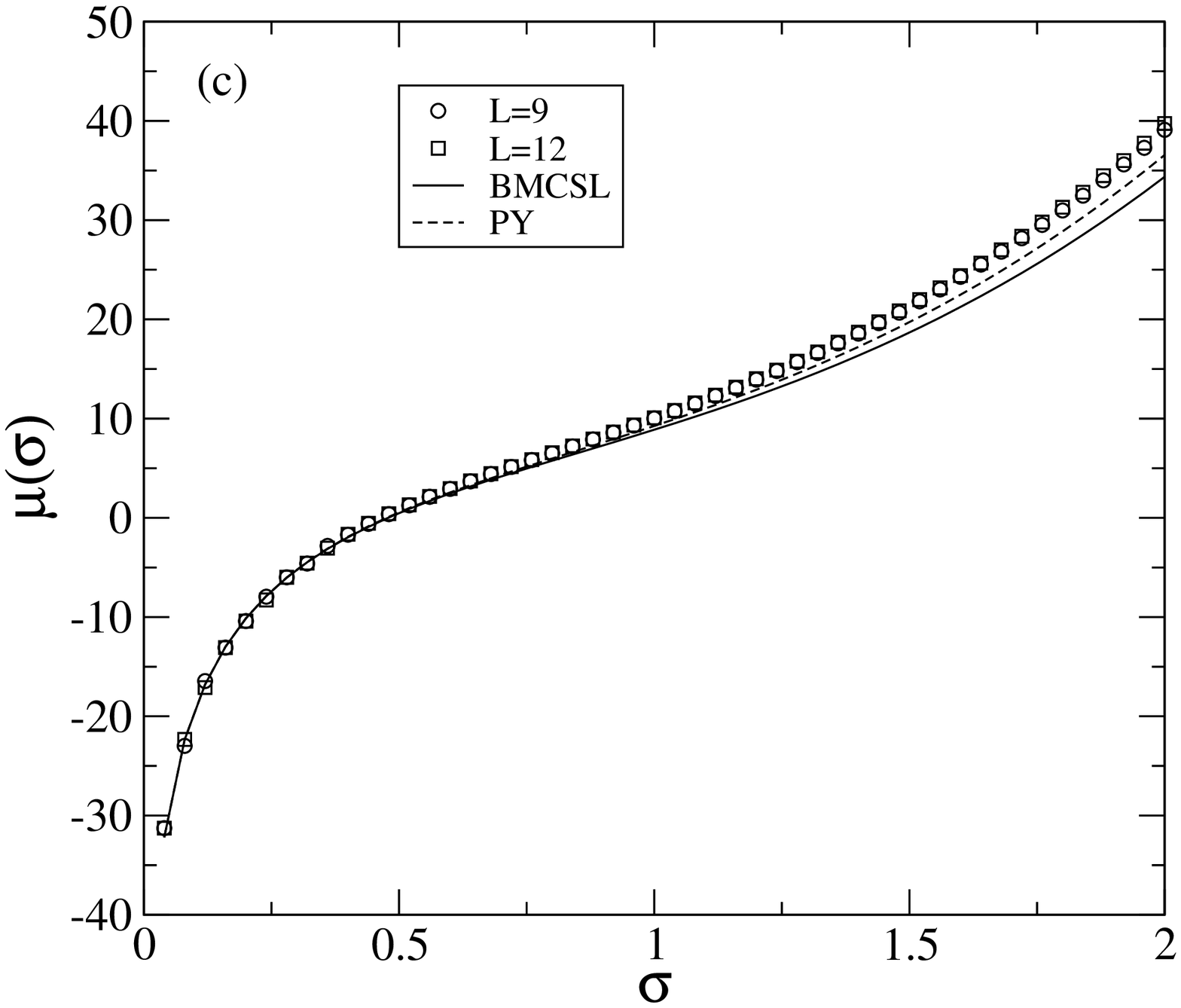}}} 
\vspace*{1cm}

\caption{Chemical potential distribution $\mu(\sigma)$ for the Schulz
density distribution ($z=15$, $\sigma_c=3$), for system sizes $L=9$
and $L=12$. {\bf (a)} $\eta=0.056$. Here the inset shows the excess
chemical potential given by eq.~\protect\ref{eq:muex}; {\bf (b)} $\eta=0.257$;
{\bf (c)} $\eta=0.426$. Also shown in each case are the predictions of
the BMCSL and PY equations of state. Statistical errors do not exceed
the symbol sizes.}

\label{fig:sz15_comp}
\end{figure}

The results for the Schulz distribution with $z=15$ at the low volume
fraction $\eta=0.056$ are shown in fig.~\ref{fig:sz15_comp}(a). At
first sight there is good agreement between the $L=9$ and $L=12$
simulation data and both the BMCSL and PY equations of state over the
entire region of $\sigma$. However, closer inspection reveals
appreciable discrepancies between theory and simulation, not visible on
the scale of $\mu(\sigma)$. These are apparent if one suppresses the
dominant ideal gas contribution to expose the {\em excess} chemical
potential, given by (see appendix~\ref{sec:append})
\begin{equation}
\mu_{\rm ex}(\sigma)=\mu(\sigma)-\ln[\rho_0n(\sigma)]\:.
\label{eq:muex}
\end{equation}
This quantity is plotted in the inset of fig.~\ref{fig:sz15_comp}(a),
from which one sees that compared to the simulation results, the EOS
slightly underestimate $\mu(\sigma)$. 

Fig.~\ref{fig:sz15_comp}(b) shows the results for the $z=15$ Schulz
distribution at the moderate volume fraction $\eta=0.257$. Again there
is good agreement between the $L=9$ and $L=12$ simulation data
suggesting that finite-size effects are insignificant. Here, however,
discrepancies between the BMCSL and PY equations of state are larger
than at the lower value of $\eta$, being evident on the scale of the
{\em absolute} chemical potential. One sees that both EOS significantly
underestimate $\mu(\sigma)$ for large $\sigma$, although they agree
quite well with one another. A similarly high level of agreement between
the data from the $L=9$ and $L=12$ system sizes is manifest at the
higher volume fraction $\eta=0.426$, fig.~\ref{fig:sz15_comp}(c). Here
the PY equation of state is seen to fare somewhat better than the BMCSL
equation although both underestimate $\mu(\sigma)$ substantially towards
the upper end of the $\sigma$ range.

\begin{figure}[h]
\setlength{\epsfxsize}{7.0cm}
\centerline{\mbox{\epsffile{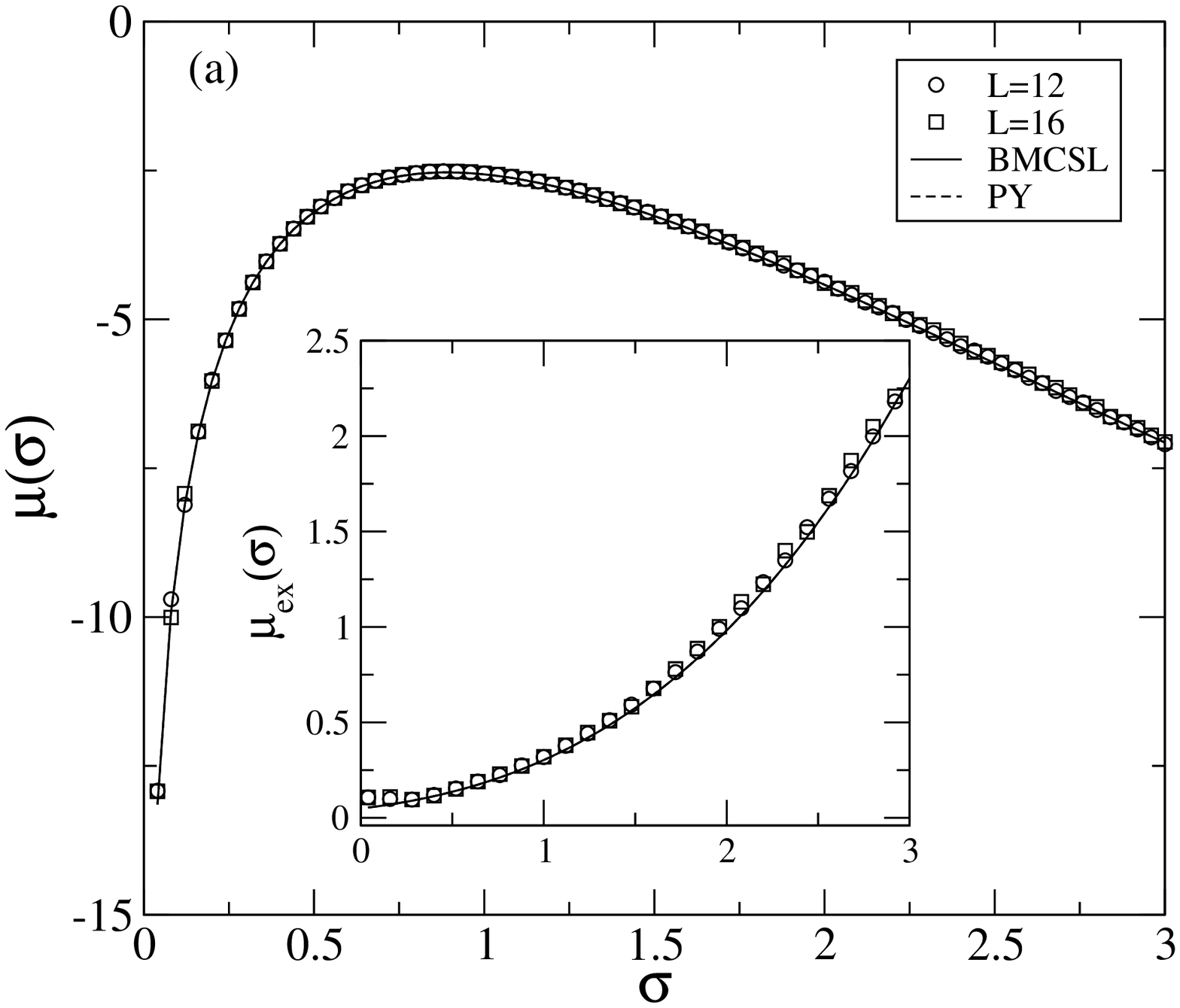}}} 
\setlength{\epsfxsize}{7.0cm}
\vspace*{0.5cm}
\centerline{\mbox{\epsffile{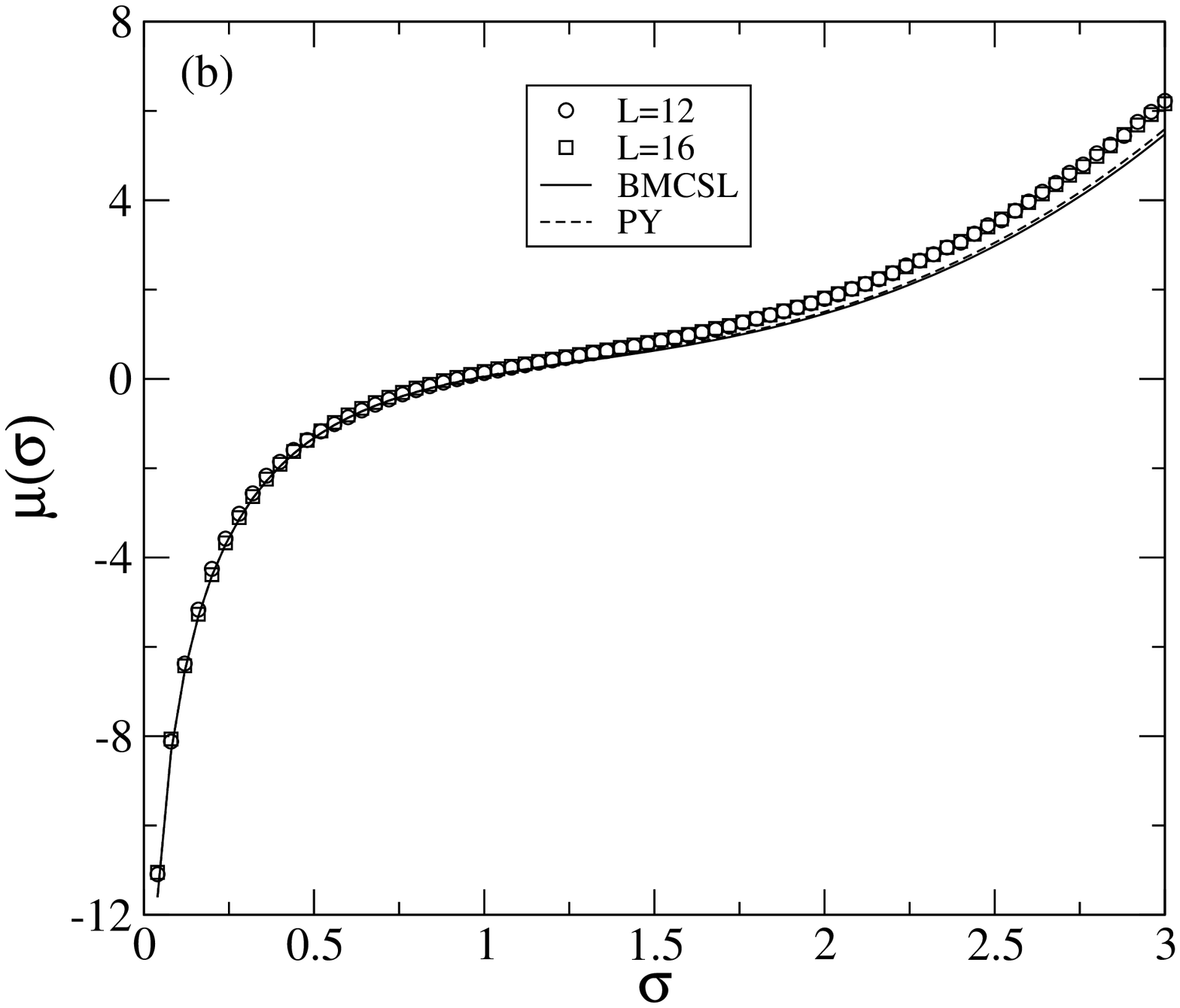}}} 
\setlength{\epsfxsize}{7.0cm}
\vspace*{0.5cm}
\centerline{\mbox{\epsffile{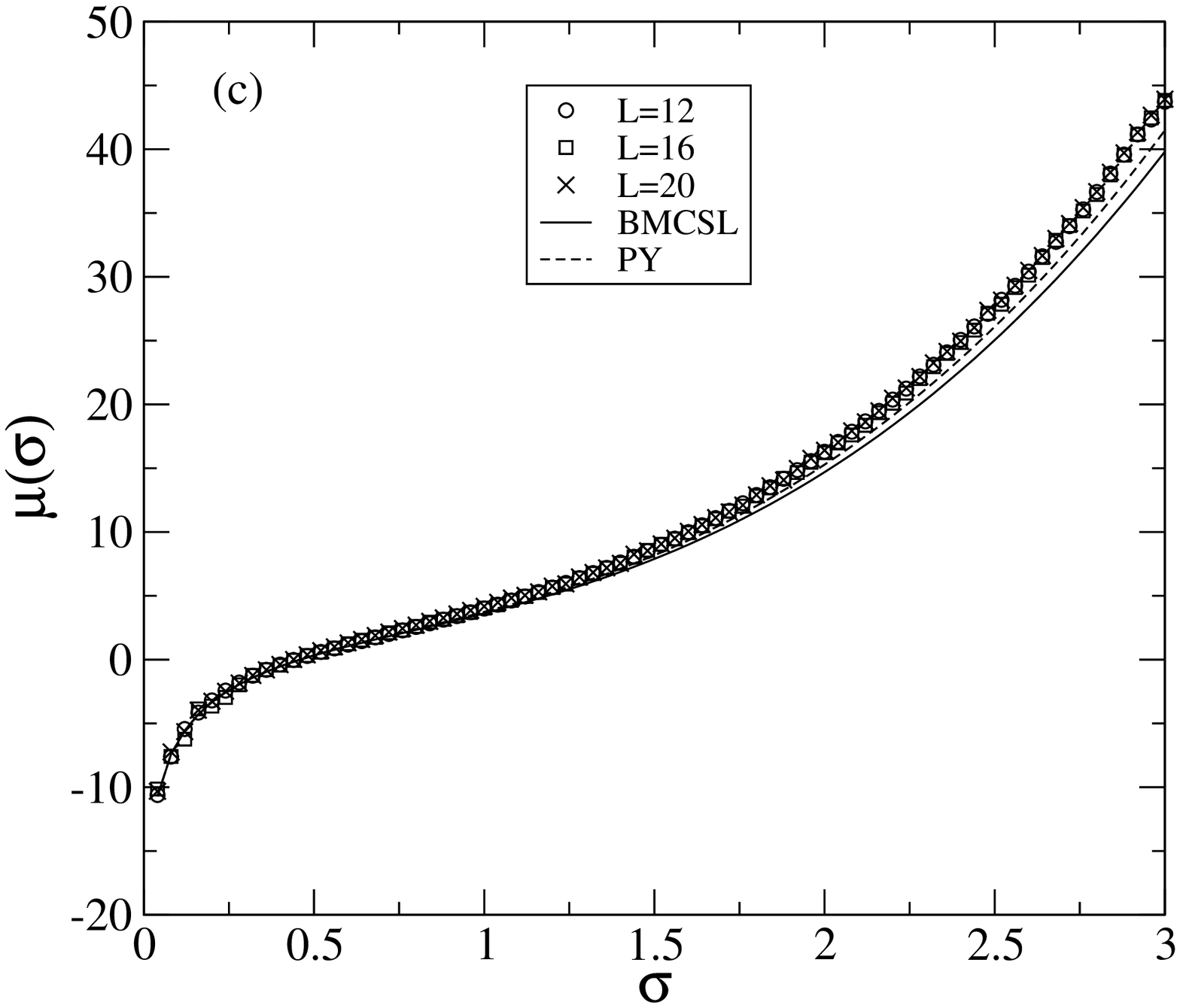}}} 
\vspace*{1cm}

\caption{Chemical potential distribution $\mu(\sigma)$ for the Schulz
density distribution ($z=5$, $\sigma_c=4$). {\bf (a)} $\eta=0.048$,
$L=12, 16$, the inset shows the excess chemical potential given by
eq.~\protect\ref{eq:muex}; {\bf (b)} $\eta=0.231$, $L=12, 16$; {\bf (c)}
$\eta=0.36$ $L=12, 16, 20$. Also shown in each case are the predictions
of the BMCSL and PY equations of state. Statistical errors do not
exceed the symbol sizes.}

\label{fig:sz5_comp}
\end{figure}

A similar picture emerges for the Schulz distribution with $z=5$
(fig.~\ref{fig:sz5_comp}). Again both equations of state underestimate
$\mu(\sigma)$, even at the lowest volume fraction, although the PY
equation corresponds significantly more closely to the simulation
results at high volume fractions than does the BMCSL equation. We could
again discern no evidence for appreciable finite-size effects in the
simulation results.

\begin{figure}
\setlength{\epsfxsize}{8.0cm}
\vspace*{1cm}
\centerline{\mbox{\epsffile{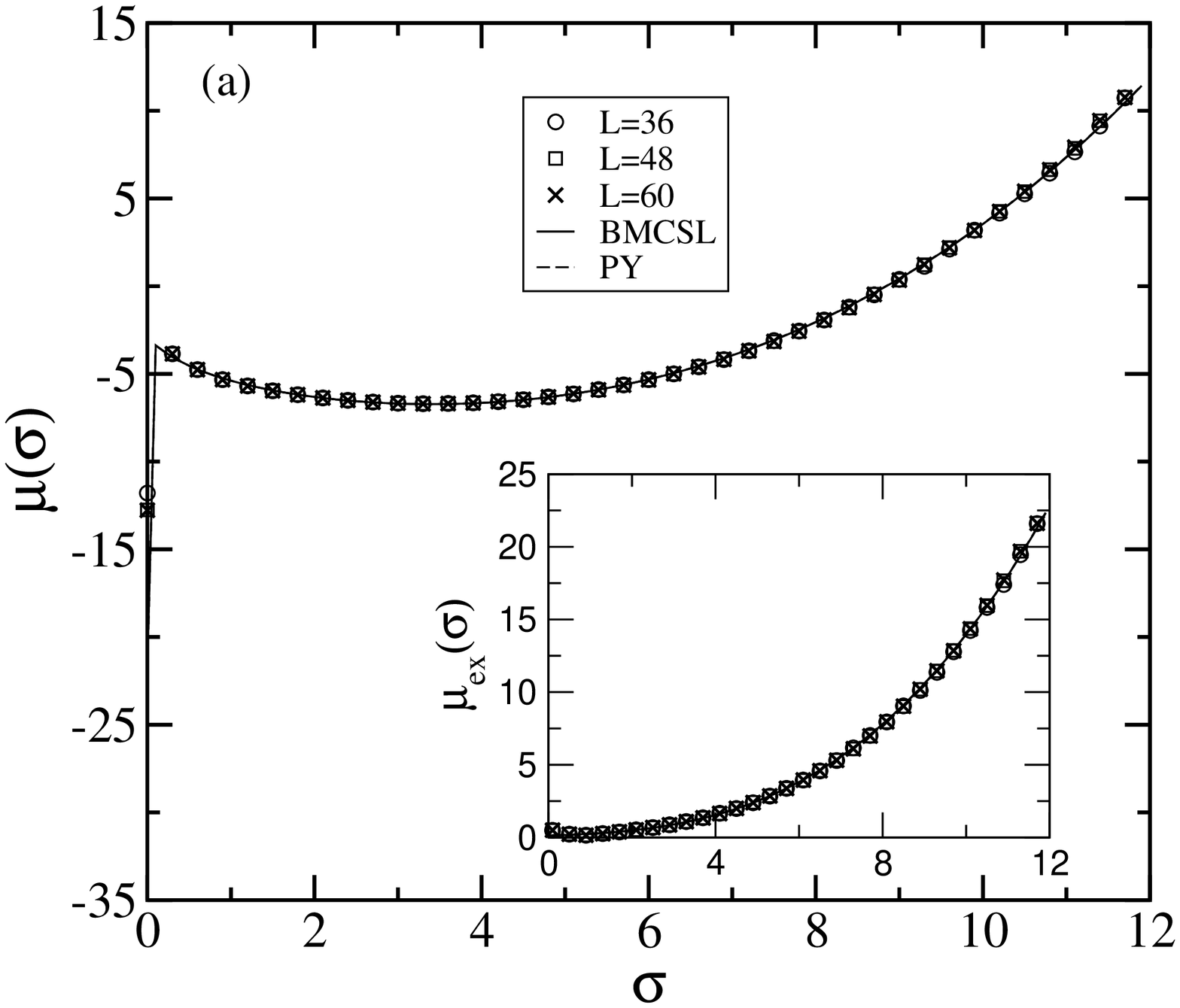}}} 
\setlength{\epsfxsize}{8.0cm}
\centerline{\mbox{\epsffile{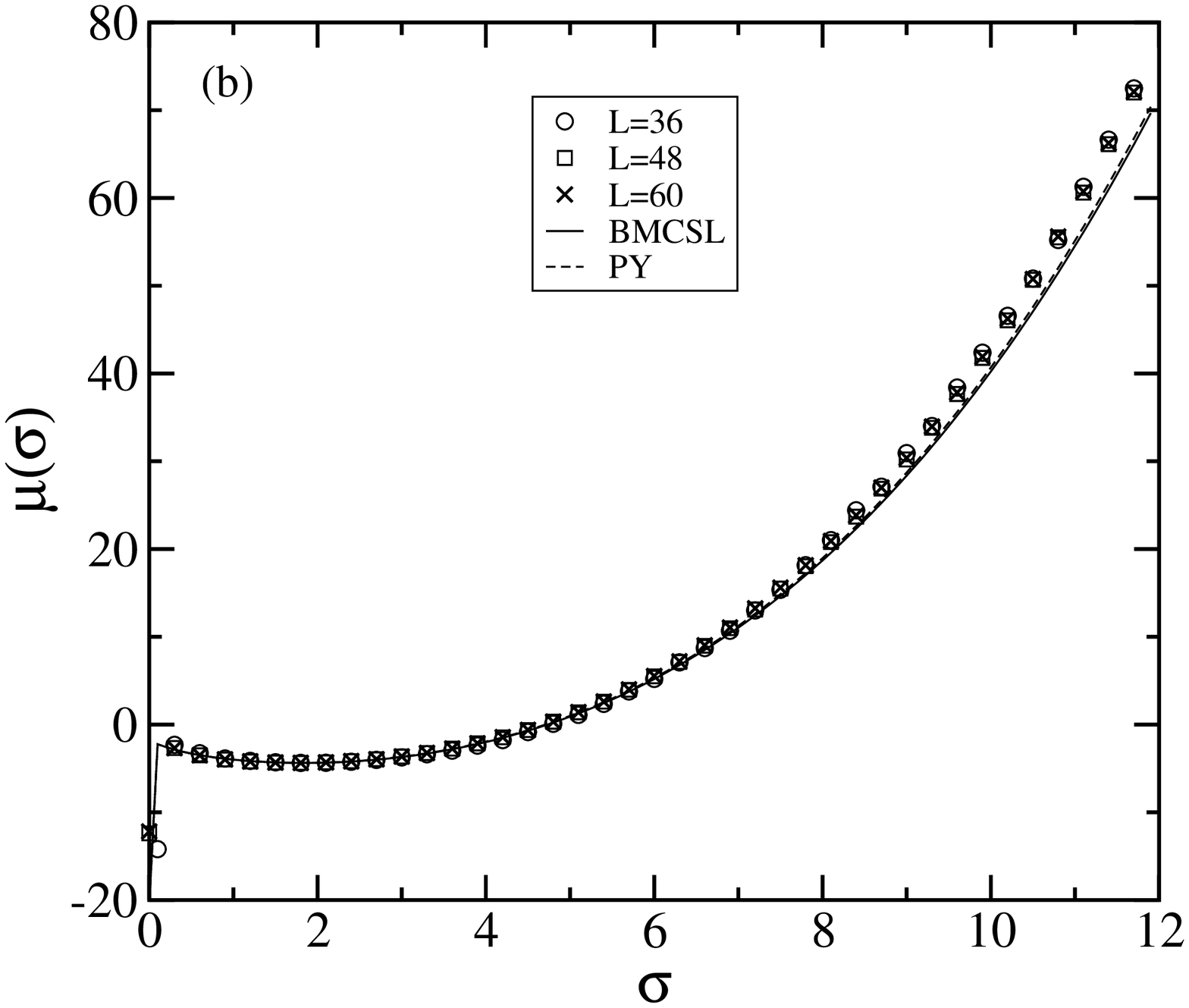}}} 
\vspace*{1cm}

\caption{Chemical potential distribution $\mu(\sigma)$ for the
log-normal density distribution ($W=2.5$, $\sigma_c=12$) for system
sizes $L=36, 48, 60$. {\bf (a)} $\eta=0.126$, the inset shows the
excess chemical potential given by eq.~\protect\ref{eq:muex}; {\bf (b)}
$\eta=0.307$. Also shown in both cases are the predictions of the BMCSL
and PY equations of state. Statistical errors do not exceed the symbol
sizes.}

\label{fig:ln_comp}
\end{figure}

In seeking to compare the results for the log-normal system with the
EOS predictions it is essential to bear in mind the importance of
truncation effects. The moments of a truncated log-normal distribution
can differ dramatically from those of the full distribution even for
large values of $\sigma_c$. In order to facilitate a fair comparison
with theory, the analytic form of $\mu(\sigma)$ must therefore be
calculated using the moments of the {\em same} truncated distribution
as employed in the simulations. The results of performing this
comparison are presented in fig.~\ref{fig:ln_comp}. At low volume
fraction, there is good agreement between the EOS predictions and the
simulation results, while at high volume fraction the EOS underestimate
the measured $\mu(\sigma)$ considerably, with the degree of discrepancy
increasing towards the tail of the distribution. Once again we could
discern no evidence of finite-size effects within the statistical
uncertainties of our data.

\begin{figure}
\setlength{\epsfxsize}{7.0cm}
\centerline{\mbox{\epsffile{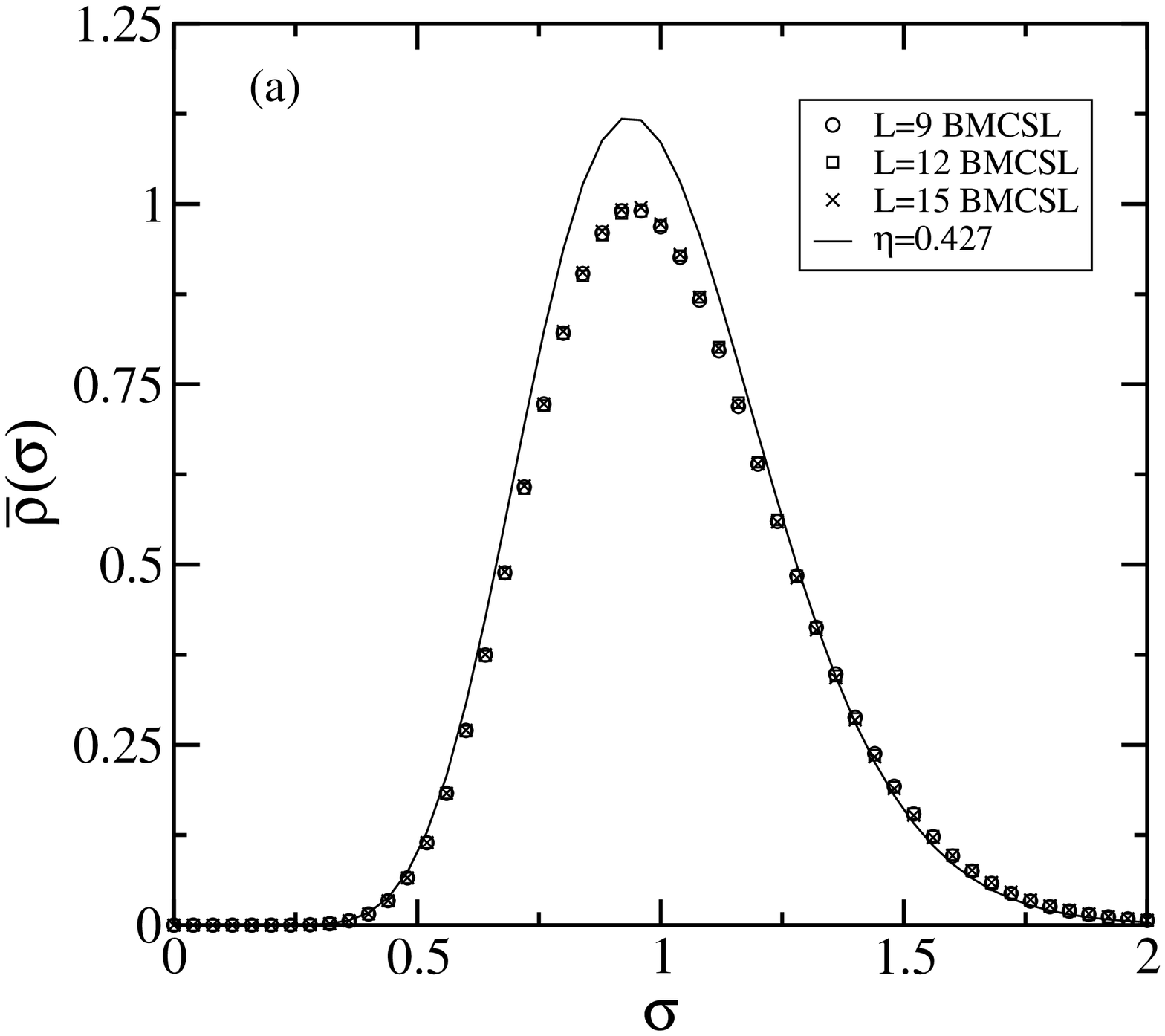}}} 
\vspace*{0.3cm}
\setlength{\epsfxsize}{7.0cm}
\centerline{\mbox{\epsffile{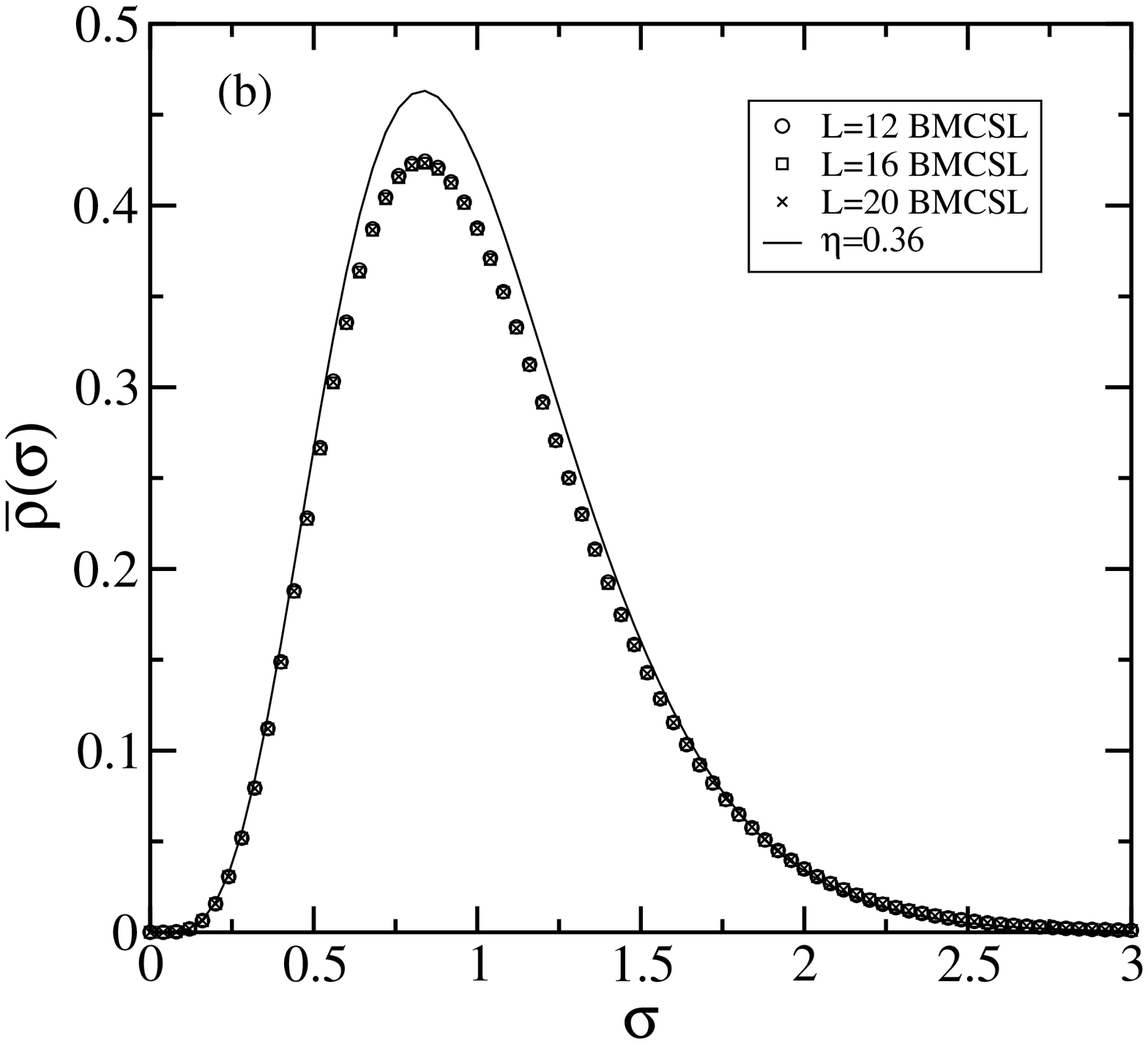}}} 
\vspace*{0.3cm}
\setlength{\epsfxsize}{7.0cm}
\centerline{\mbox{\epsffile{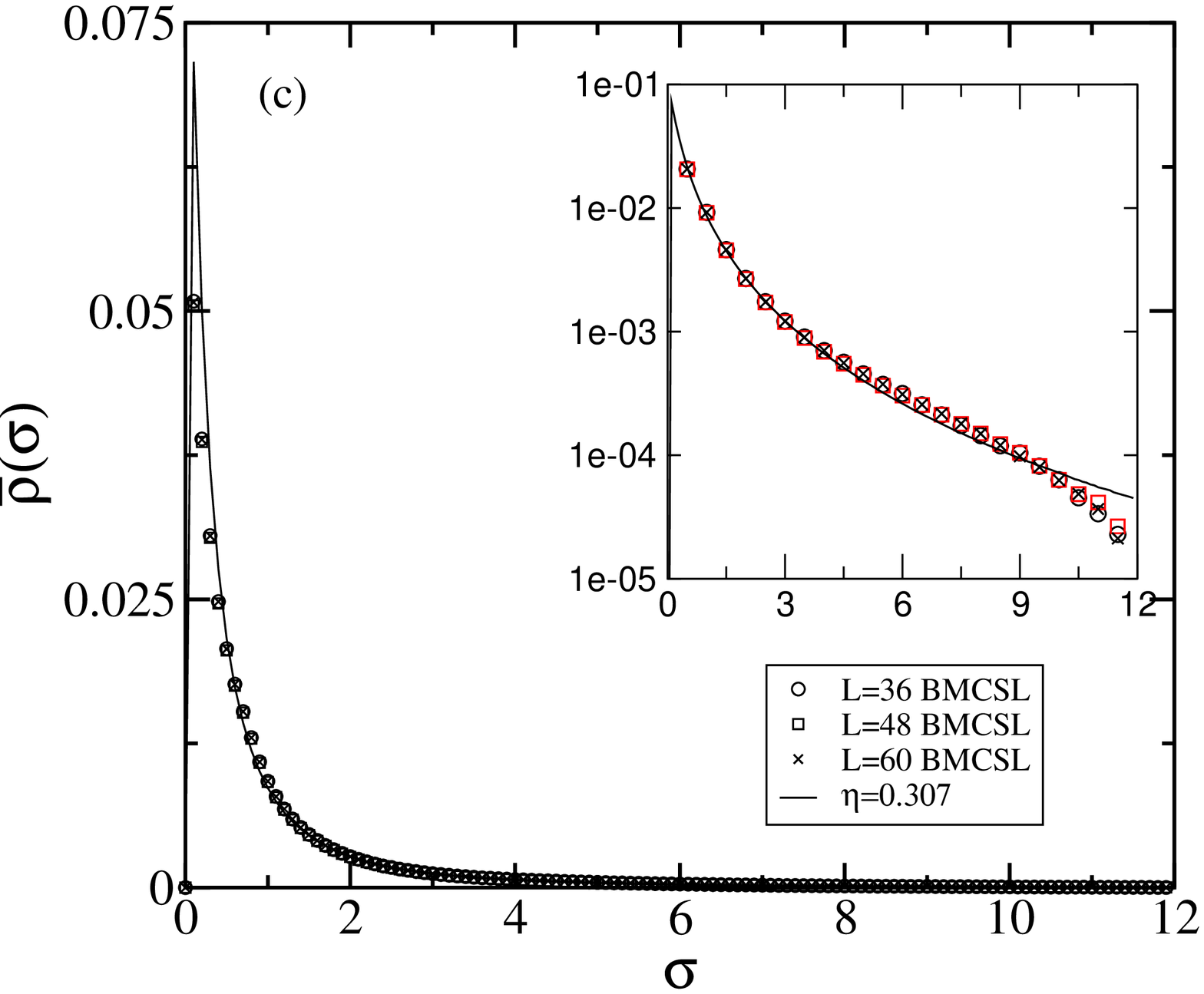}}} 
\vspace*{0.5cm}

\caption{The measured density distribution $\rho(\sigma)$ obtained
using forms of $\mu(\sigma)$ predicted by the BMCSL equation of state
at the stated volume fraction. {\bf (a)} Schulz ($z=15$), $\eta=0.427$,
$L=9, 12, 15$; {\bf (b)} Schulz ($z=5$), $\eta=0.36$, $L=12, 16, 20$;
{\bf (c)}  log-normal distribution,  $\eta=0.307$, $L=36, 48, 60$, the
inset shows the same data on a log scale.  In each case the density
distribution from which $\mu(\sigma)$ derives is shown as a solid line.
Statistical errors do not exceed the symbol sizes.} 

\label{fig:fse} 
\end{figure}

The results presented above indicate that the BMCSL and PY equations
fail to provide a quantitatively accurate description of the chemical
potential distribution, particularly when the volume fraction is
large. It is instructive to examine the implications of this finding
for calculations of {\em densities}, which depends very sensitively
(indeed exponentially so) on the chemical potential. To this end we have
studied the degree to which the form of $\mu(\sigma)$ calculated via
the BMCSL equation from some prescribed $\rho(\sigma)$, actually yields
this density distribution when input to a simulation. The results of
performing this comparison are shown in figures~\ref{fig:fse}(a) --
\ref{fig:fse}(c) for each of the three fluids at a high volume
fraction. In each instance, the solid line shows the input density
distribution $\rho(\sigma)$ from which $\mu(\sigma)$ is calculated. The
data points show the simulation results obtained using this form of
$\mu(\sigma)$ for $3$ different system sizes. As the figures clearly
demonstrate, the measured form of $\rho(\sigma)$ deviate substantially
from the prediction. 

\begin{figure}
\setlength{\epsfxsize}{8.0cm}
\vspace*{1cm}
\centerline{\mbox{\epsffile{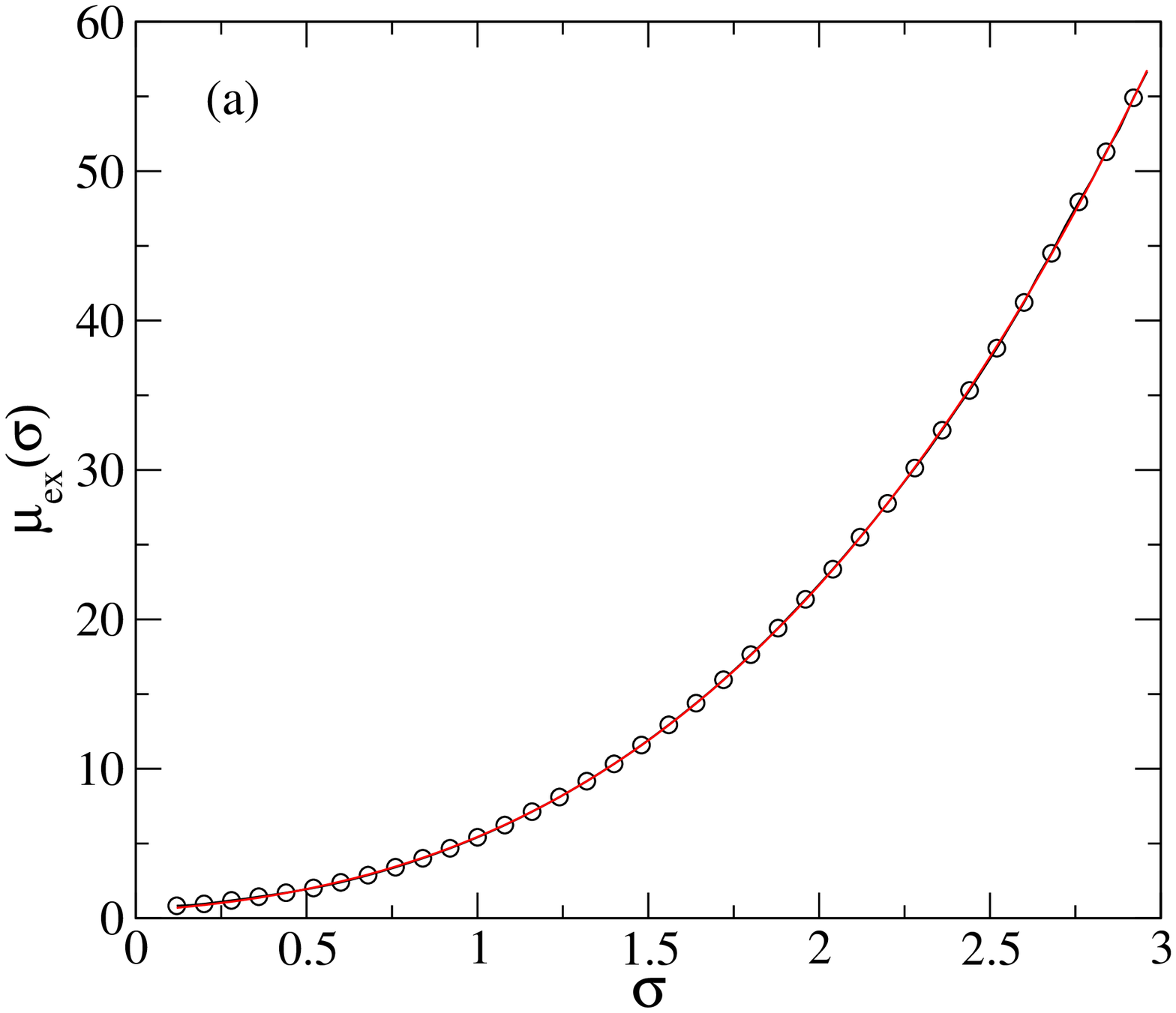}}} 
\setlength{\epsfxsize}{8.0cm}
\centerline{\mbox{\epsffile{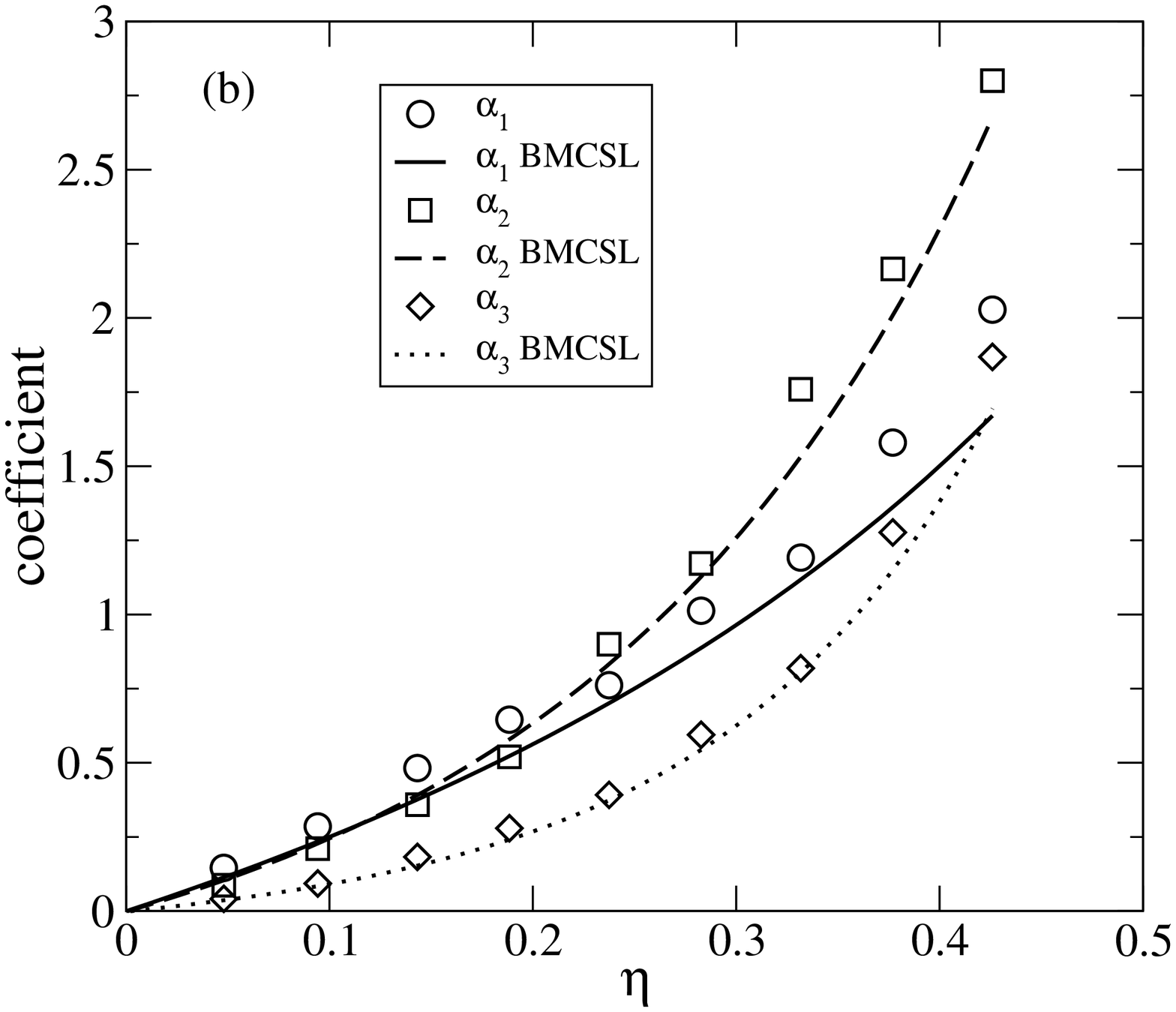}}} 
\vspace*{1cm}

\caption{ {\bf (a)} Data points show the measured excess chemical
potential distribution $\mu_{\rm ex}(\sigma)$ for the Schulz density
distribution ($z=5$) at $\eta=0.377$, for system size $L=12$. The solid
lines shows a fit to the data of the form given by
eq.~\protect\ref{eq:muexfit}. {\bf (b)} Values of the fit coefficients
$\alpha_1,\alpha_2,\alpha_3$ for a selection of values of the volume
fraction $\eta$. The solid lines show the predictions of the BMCSL equation.
Statistical errors do not exceed the symbol sizes.}

\label{fig:coeffs}
\end{figure}

Finally in this section, we examine the moment structure of the excess
chemical potential $\mu_{\rm ex}(\sigma)$ given by eq.~\ref{eq:muex}. Both
the BMCSL and PY equations assume that $\mu_{\rm ex}(\sigma)$ is
expressible in terms of a cubic polynomial in $\sigma$. Bartlett has
reached the same conclusion using geometrical arguments inspired by
scaled particle theory \cite{BARTLETT97}. We have investigated this
proposal by fitting our data to the expression

\begin{equation}
\mu_{\rm ex}(\sigma)=-\ln(1-\eta)+\alpha_1\sigma+\alpha_2\sigma^2+\alpha_3\sigma^3\:,
\label{eq:muexfit}
\end{equation} 
where the constant term is fixed by the requirement that in the limit
$\sigma\to 0$ the probability of inserting a sphere is proportional to
$1-\eta$. We find that all our data are fitted very well by this
expression; fig.~\ref{fig:coeffs}(a) shows a typical fit for the case
of the Schulz fluid with $z=5$ at $\eta=0.377$. Also shown,
(fig.~\ref{fig:coeffs}(b)) are the fit coefficients $\alpha_1,
\alpha_2, \alpha_3$ for various values of $\eta$, together with the
predictions of the BMCSL equation of state. One sees that at high
volume fraction, the BMCSL underestimates all coefficients, the
relative discrepancy being marginally larger for the $\alpha_1$
coefficient than for the others.

\section{Discussion and conclusions}
\label{sec:conclusions}

In summary, we have presented a grand canonical simulation method for
studying polydisperse fluids. The method utilizes histogram
extrapolation techniques to track efficiently an arbitrary path of
interest through the space of the density distribution $\rho(\sigma)$. 
We have applied it to the specific problem of obtaining the dilution
line properties of size-disperse hard spheres. It should also prove
useful in studying more general species of polydisperse fluids and their
phase transitions, both in the bulk and confined geometries. We intend
to report on such extensions in future communications.

Previous simulation studies of polydisperse fluids have generally
operated within a semi-grand canonical framework \cite{KOFKE} in which a
fixed number of particles are studied either at constant pressure (see
eg. ref.~\cite{BOLHUIS}) or within a Gibbs Ensemble MC scheme
\cite{KRISTOF}. In common with the present work, these studies utilized
a fluctuating particle size distribution, realized by means of MC
resizing moves controlled by a chemical potential distribution.  In
contrast to our approach, however, thermodynamic properties were studied
as a function of the shape of the activity distribution
$\exp(\mu(\sigma))$; no constraints were placed on the conjugate density
distribution, which was consequently free to adopt whichever functional
form minimized the free energy for the imposed $\mu(\sigma)$. In view of
this, one might question the extent to which the simulation results
reflect the actual situation in realistic systems. One situation in
which the lack of a constrained density distribution might be relevant
is the interesting issue of the influence of polydispersity on the
freezing of hard spheres \cite{BARTLETT99}. This was investigated in
ref.~\cite{BOLHUIS}, using semi-grand canonical MC and Gibbs-Duhem
integration. It would be of considerable interest to harness the
approach described in the present work to investigate the
effects on the freezing behaviour of varying the width of the density
distribution whilst constraining its shape to some physically realistic
form.

In the present study, attention was focused on fluids having a high
degree of polydispersity. The motivation for this choice was two-fold.
First, models exhibiting a wide density distribution provide a suitably
testing challenge against which to assess the effectiveness of our
method. Second, there exist in the literature a number of interesting
predictions concerning the role of depletion forces in highly
polydisperse systems. For instance, it has been suggested by several
authors that attractive depletion forces might engender novel phase
transitions in polydisperse hard spheres \cite{WARREN,CUESTA,SEAR}.
Specifically, Sear has suggested that a fluid of hard spheres having a
log-normal size distribution will be unstable with respect to
crystallization of the large particles at {\em all} finite volume
fractions \cite{SEAR}.  By contrast, Cuesta \cite{CUESTA} has predicted
that a sufficiently wide log-normal distribution will exhibit
fluid-fluid phase separation at some finite density. It seems likely
that for a truncated size distribution a phase transition will not
occur for arbitrarily small volume fraction because of the absence of
the largest particles which mediate the greatest depletion forces. Thus
our simulation results for the truncated distribution are able neither
to conclusively confirm nor refute these predictions. It suffices to
say that we observed no evidence of crystallization in the particular
truncated log-normal fluid studied, up to a volume fraction of
$\eta=0.33$. Similarly, within the range of accessible volume
fractions, no evidence for phase transitions was observed in either of
the two Schulz fluids studied.

Turning finally to the comparison between our simulation results and
the predictions of the polydisperse equations of state, we find that
neither the BMCSL nor PY equations offer a quantitatively accurate
description of the thermodynamics of hard spheres for large
polydispersity and at high volume fraction \cite{TAMARA}.  Both
equations underestimate $\mu(\sigma)$ at all fluid densities over the
entire range of $\sigma$, implying that they underestimate the
Helmholtz free energy density $f=\int_0^{\rho_0} \mu(\sigma |
\rho_0^\prime n(\sigma)) d\sigma d\rho_0^\prime$ and hence overestimate the
stability of the fluid. The magnitude of this overestimate becomes
more pronounced the greater the volume fraction. Interestingly we find
that in this regime the PY equation performs appreciably better than
the BMCSL equation despite the fact (cf. appendix~\ref{sec:append})
that the latter derives from a monodisperse hard sphere equation of
state which has been found to be superior to the PY approximation.  It
remains to be seen to what extent the overestimate of fluid stability
impinges on the results of existing calculations of depletion-force
induced phase separation based on the BMCSL and PY approximations
\cite{WARREN,CUESTA}. In any case, our results should provide a useful
testing ground for any future improvements to the existing
polydisperse equations of state \cite{BARRIO}.

\acknowledgements

NBW acknowledges helpful discussions with Marcus M\"{u}ller. This work
was supported by The University of Liverpool Research Development Fund
and the UK CCP5 group. PS acknowledges financial support through EPSRC
grant GR/R52121.

\newpage

\newpage

\newpage

\newpage

\newpage

\newpage 
\newpage

\newpage

\appendix
\section{Polydisperse hard sphere equations of state}
\label{sec:append}

The equations of state that we quote here express the chemical
potential as a function the sphere diameter $\sigma$. The expression
due to Salacuse and Stell \cite{SALACUSE} is a generalization of the
Percus-Yevick result for monodisperse hard spheres:
\begin{eqnarray}
\mu_{\rm PY}(\sigma)&=&\ln[\rho_0 n(\sigma)]-\ln(1-\zeta^3)+
\frac{\sigma^3\zeta_0+3\sigma^2\zeta_1+3\sigma\zeta_2}{1-\zeta_3}
\nonumber\\
&\:&
+\frac{3\sigma^3\zeta_1\zeta_2+9\sigma^2\zeta_2^2/2}{(1-\zeta_3)^2}
+\frac{3\sigma^3\zeta_2^3}{(1-\zeta_3)^3}
\end{eqnarray}
The BMCSL equation of state \cite{BOUBLIK,MANSOORI}, on the other hand,
generalizes the Carnahan-Starling expression, which for
monodisperse hard spheres is more accurate than the Percus-Yevick
result. It is given by
\begin{eqnarray}
\mu_{\rm BMCSL}(\sigma) &=& \ln[\rho_0n(\sigma)]
+\left(3\sigma^2\zeta_2^2/\zeta_3^2-2\sigma^3\zeta_2^3/\zeta_3^3\right)
\ln(1-\zeta^3)
\nonumber\\
&\:&
+\frac{\sigma^3(\zeta_0-\zeta_2^3/\zeta_3^2)+3\sigma^2\zeta_1+3\sigma\zeta_2}
{1-\zeta_3}
\nonumber\\
&\:&+\frac{\sigma^3(3\zeta_1\zeta_2-\zeta_2^3/\zeta_3^2) + 3
\sigma^2\zeta_2^2/\zeta_3}
{(1-\zeta_3)^2}
+\frac{2\sigma^3\zeta_2^3}{\zeta_3(1-\zeta_3)^3}
\end{eqnarray}
where $\zeta_n=\pi\rho_0m_n/6$ and $m_n$ is the $n$-th moment of
$\overline{\rho}(\sigma)$; note that $\zeta_3=\eta$, the volume
fraction of hard spheres.

Both equations of state include the ideal contribution to the chemical
potential, $\mu_{\rm ideal}(\sigma)=\ln[\rho_0n(\sigma)]$. For general
temperature --- remember that we set $\beta=1$ for our hard sphere
system --- this contribution would read $\beta\mu_{\rm ideal}(\sigma) =
\ln\bar\rho(\sigma)$. Because there has been some discussion in the
literature regarding how the ideal chemical potential for polydisperse
systems should be assigned (see e.g.~\cite{SALACUSE}), it may be helpful
to note that, within the grand canonical framework, $\mu_{\rm
ideal}(\sigma)$ is unambiguously defined via the grand canonical
partition function. Indeed, for an ideal system (${\cal H}_N=0$), one
readily finds from eq.~\ref{eq:bigzdef} that $\bar\rho(\sigma)=
\exp(\beta\mu(\sigma))$, giving $\beta\mu_{\rm
ideal}(\sigma)=\ln\bar\rho(\sigma)$ as before. One may worry about
dimensions in the argument of the log here; but the definition in
eq.~\ref{eq:bigzdef} already implies that dimensionless units are used
for lengths and for $\sigma$. Were this not the case, the integrals over
$\vec{r}_i$ and $\sigma_i$ would need to be normalized by a unit volume
$v_0$ and a unit value $\sigma_0$ for the polydisperse attribute, and
the ideal chemical potential would read $\beta\mu_{\rm ideal}(\sigma) =
\ln[v_0\sigma_0\bar\rho(\sigma)]$, with the argument of the log now
manifestly dimensionless. The normalization constant $v_0\sigma_0$ could
in fact be made dependent on $\sigma$; this would just give a
$\sigma$-dependent shift in the zero of the chemical potential scale.

\newpage


\begin{thebibliography}{99}
\bibitem{FOOT1} Superior to the CE is a constant-(NpT) ensemble, in
which the overall density can fluctuate via changes in the system
volume. However the problems associated with a fixed realization of
$N(\sigma)$ remain.
\bibitem{FOOT2} Conceptually and operationally (cf.
sec.~\ref{sec:implement}) it is expedient to regard $\rho(\sigma)$ as a
binned distribution (histogram) in the limit in which the bin size
$\Delta\sigma$ goes to zero. The product in eq.~\ref{eq:jointdef} is to
be understood to run over all bins.
\bibitem{NOTE0} There is some license with regards to the choice of
values for $\sigma_c$ and $M$. Clearly, where computationally
practical, $\sigma_c$ should be chosen to exceed the maximum value of
$\sigma$ for which significant weight occurs in $\rho(\sigma)$ at any
point along the phase space path of interest. The choice of $M$ should
strike a suitable balance between data resolution and smoothness of the
histogram.
\bibitem{NOTE0A}  For anything other than hard particles, the energy
$u$ should also be written to the list, because it is required for
histogram reweighting.
\bibitem{FOOT5} Once the extrapolation has been made, and a new
simulation performed, it is advisable to check the reliability of the
extrapolation by comparing the measured $\rho(\sigma)$ with the target
function.
\bibitem{betanote} Owing to the lack of a temperature scale in hard
sphere systems, we implicitly adopt the convention of assigning $\beta=1$.
\bibitem{FOOT6} For the case of hard spheres, the form for $\mu(\sigma)$
could equally well have been bootstrapped at low densities using the
predictions of one of the analytical equations of state
\protect\cite{BOUBLIK,MANSOORI,SALACUSE}.
\end{thebibliography}
\end{document}